\documentclass[a4paper,fleqn,usenatbib]{mnras}
\usepackage[normalem]{ulem}

\usepackage{mathptmx}

\usepackage[T1]{fontenc}
\usepackage{ae,aecompl}
\usepackage[english]{babel}
\usepackage[utf8]{inputenc}

\usepackage{graphicx}	
\usepackage{amsmath}	
\usepackage{amssymb}	

\usepackage{longtable}
\usepackage{lipsum}
\usepackage{txfonts}
\usepackage{threeparttable}
\usepackage[bottom]{footmisc}

\title[New Insights into Time Series Analysis IV]{\centering New Insights into
Time Series Analysis IV: \\ Panchromatic and Flux Independent Period
Finding Methods}

\author[C. E. Ferreira Lopes et al.]{
C. E. Ferreira Lopes$^{1}$\thanks{E-mail: ferreiralopes1011@gmail.com},
N. J. G.~Cross$^2$, 
F.  Jablonski$^{1}$
\\
$^{1}$National Institute For Space Research (INPE/MCTI), Av. dos Astronautas, 1758 – São José dos Campos – SP, 12227-010, Brazil  \\
$^{2}$SUPA (Scottish Universities Physics Alliance) Wide-Field Astronomy Unit, Institute for Astronomy, School of Physics and Astronomy,\\ University of Edinburgh, Royal Observatory, Blackford Hill, Edinburgh EH9 3HJ, UK
}

\date{Accepted XXX. Received YYY; in original form ZZZ}

\pubyear{2018}

\begin{document}
\label{firstpage}
\pagerange{\pageref{firstpage}--\pageref{lastpage}}
\maketitle

\begin{abstract}
New time-series analysis tools are needed in disciplines as diverse as
astronomy, economics and meteorology.
In particular, the increasing rate of data collection at multiple wavelengths requires new approaches
able to handle these data. 
The panchromatic correlated indices $K^{(s)}_{(fi)}$ and $L^{(s)}_{(pfc)}$ are adapted to quantify the smoothness of a phased light-curve resulting in new period-finding methods applicable to single- and multi-band data. Simulations and observational data are used to test our approach. The results were used to establish an analytical equation for the amplitude of the noise in the periodogram for different false alarm probability values, to determine the dependency on the signal-to-noise ratio, and to calculate the yield-rate for the different methods.  The proposed method has similar efficiency to that found for the String Length  period method. The effectiveness of the panchromatic and flux independent period finding methods in single waveband as well as multiple-wavebands that share a fundamental frequency is also demonstrated in real and simulated data.
\end{abstract}

\begin{keywords}
methods: data analysis -- methods: statistical -- techniques: photometric -- astronomical databases: miscellaneous -- stars: variables: general
\end{keywords}

\section{Introduction}\label{introduction}	

If the brightness variations of a variable star are periodic, one can fold
the sparsely sampled light-curve with that period and inspect the magnitude as
a function of phase plot. This will be equivalent to all the measurements of
the star brightness taken within one period. The shape of the phased light-curve and the period allow one to determine the physical nature of variability (pulsations, eclipses, stellar activity, etc.). If the light-curve is folded with a wrong period, the magnitude measurements will be all over the place rather than align into a smoothly varying function of the phase. Other methods figure out the best period fitting a specific model into the phased light curve, like a sine function. The most common methods used in astronomy are the following:  the Deeming method \citep[][]{Deeming-1975}, phase dispersion minimization  \citep[PDM -][]{Stellingwerf-1978, Dupuy-1985}, string length minimization \citep[SLM  -][]{Lafler-1965,Dworetsky-1983,Stetson-1996,Clarke-2002}, information entropy \citep[][]{Cincotta-1995}, the analysis of variance \citep[ANOVA -][]{Schwarzenberg-Czerny-1996}, and the Lomb-Scargle periodogram  and its extension using error bars \citep[LS and LSG -][]{Lomb-1976,Scargle-1982,Zechmeister-2009}. All of these methods require as input the minimum frequency ($f_{min}$), the maximum frequency ($f_{max}$), and the sampling frequency (or the number of frequencies tested - $N_{freq}$). The input parameters and their constraints to determine reliable variability detections were addressed by \citet[][]{FerreiraLopes-2018papIII},  where a summary of recommendations on how to determine the sampling frequency and the characteristic period and amplitude of the detected variations is provided. From these constraints, a good period finding method should find all periodic features if the time series has enough measurements covering nearly all variability phases \citep[][]{Carmo2020}.

Light curve shape, non-Gaussianity of noise, non-uniformities in the data spacing, and multiple periodicities modify the significance of the periodogram and to increase completeness and reliability, more than one period finding method is usually applied to the data \citep[e.g.][]{Angeloni-2012,FerreiraLopes-2015wfcam,FerreiraLopes-2015cycles}. The capability to identify the "true" period is increased by using several methods (see Sect. \ref{sec_pcaution}). However, this does not prevent the appearance of spurious results. Therefore, new insights into signal detection which provide more reliable results are welcome mainly when the methods provide dissimilar periods.  Moreover, the challenge of big-data analysis would benefit a lot from a single and reliable detection and characterization method. The present paper is part of a series of studies performed in the project called New Insight into Time Series Analysis (NITSA), where all steps to mining photometric data on variable stars are being reviewed. The selection criteria were reviewed and improved \citep[][]{FerreiraLopes-2016papI,FerreiraLopes-2017papII}, optimized parameters to search and analyse periodic signals were introduced \citep[][]{FerreiraLopes-2018papIII}, and now new frequency finding methods are proposed to increase our inventory of tools to create and optimize automatic procedures to analyse photometric surveys. The outcome of this project is crucial if we are to efficiently select the most complete and reliable sets of variable stars in surveys like the VISTA Variables in the Via Lactea \citep[VVV - ][]{Minniti-2010,Angeloni-2014vvv}, Gaia \citep[][]{Perryman-2005}, the Transiting Exoplanet Survey Satellite \citep[TESS - ][]{Ricker-2015}, the Panoramic Survey Telescope and Rapid Response System \citep[Pan-STARRS - ][]{Chambers-2016}, a high-cadence All-sky Survey System \citep[ATLAS - ][]{Tonry-2018}, Zwicky Transient Facility \citep[ZTF - ][]{Bellm-2019} as well as the next generation of surveys like PLAnetary Transits and Oscillation of stars \citep[PLATO - ][]{Rauer-2014}  and Large Synoptic Survey Telescope \citep[LSST - ][]{Ivezic-2008}.

Many efforts are being performed to generalize for multi-band data the period finding methods. \citet[][]{Suveges-2012}  utilized the principal component analysis to optimally extract the best period using multi-band data. However, the multi-band observations must be taken simultaneously that impose an important limitation to the method. On the other hand, \citet[][]{VanderPlas-2015} introduces a general extension of the Lomb-Scargle method while \citet[][]{Mondrik-2015} for the ANOVA from single band algorithm to multiple bands not necessarily taken simultaneously. Indeed, methods combining the results from two different classes of period-determination algorithms are also being reached \citep[][]{Saha-2017}. 
 The current paper adds one piece to this puzzle.
Section \ref{sec_newmethod} describes the new set of periodic signal detection  methods as well as their limitations and constraints. Next, numerical simulations are used to test our approach in Sect. \ref{sec_simulation}. From this, the efficiency rate and the fractional fluctuation of noise (FFN) are  determined. Real data are also used to support our final results (see Sect. \ref{sec_datatest}). Finally, our conclusions are presented in Sect. \ref{sec_conclusions}.

\section{Panchromatic and flux independent frequency finding
methods}\label{sec_newmethod_intro}

The Welch-Stetson variability index \citet[][]{Stetson-1996} was generalized and new ones were performed by \citet[][]{FerreiraLopes-2016papI}. From where the panchromatic and flux independent variability indices were proposed. These indices are used to discriminate variable stars from noise. To summarise, the panchromatic index is related to the correlation amplitude (or correlation height) while the second one computes the correlation sign, i.e. if the correlation value is negative or positive without taking into account the amplitude. The flux independent index provides correlation information that is weakly dependent on the amplitude or the presence of outliers. These features enable us to reduce the misclassification rate and improve the selection criteria. Moreover, this parameter is designed to compute  correlation values among two or more observations. The correlation order ($s$), gives the number of observations correlated together, i.e., $s=2$ means correlation computed between pairs of observations and $s=3$ means that correlations are computed on triplets. However, these observations must be close in time, i.e., those observations are taken in an a interval time smaller much less than the main variability period. Inaccurate or incorrect outputs will be obtained if this restriction is not enforced. Therefore, the data sets and sources with observations close in time were named as correlated-data otherwise non-correlated-data.

The efficiency rate to detect variable stars is maximised using the panchromatic and flux independent variability indices when the number of correlations is increased, i.e when there is a strong variability between bins, but only slight differences between the measurements in each correlation bin. These variability indices only use those measurements that are close in time (i.e., a time interval much smaller than the variability period) and  hence this constraint substantially reduced the number of possible correlations for sparse data. If we consider a light-curve folded on its true variability period, with little noise, we could calculate these indices using standard correlated observations grouped in time, missing the observations where too few meet the criteria of having at least $s$ closer than $\Delta\,T$ in time. Alternatively, all measurements can be used to compute the indices if the observations are grouped by phase instead of time. It is the main idea to support the Panchromatic and flux independent frequency finding methods.

For the main variability period, the observations closed in phase should return strong correlation values. Since many variable stars show most variation as a function of phase, and little variation from period to period, recalculating the indices this way should return indices that are as strong as those grouped by time. On the other hand, if the light-curve is folded on an incorrect period and the calculated phase is no longer a useful correlation measure, so correlations will be weaker, much like
adding more noise to the data. The statistics considered in this paper are unlikely to be useful for data with multiple periodicities or if noise  keeps its autocorrelation for phased data. In the next section, we propose an  approach to compute the panchromatic and flux independent indices in phase and hence provide a new period finding method.

Be aware that, the definition of expected noise performed by
\citet[][]{FerreiraLopes-2015wfcam} needs to be corrected, as pointed by the
referee of the current paper. The authors provided the correct theoretical
definition of expected noise but the mathematical expression was incorrect. In
the case of statistically independent events, the probability that a given event
will occur is obtained by dividing the number of events of the given type by 
the total number of possible events, according to the authors. There will 
always be 2 desired permutations (either all positive or all negative) for any 
s value. However, the total number of events is $2^{s}$, not $s^{2}$ as defined
by the authors. The correct definition for expected noise value is then given by,

\begin{equation}
   P_{s} = \frac{2}{2^s} = 2^{(1-s)}
   \label{eq_newnoise}     
\end{equation}

The relative differences between the old and new definition for $s=2$
and $s=4$ are zero while for $s=3$ is $\sim11\%$. However, for s values larger
than 4 these differences increase considerably. The authors have only used the
noise definitions to set the noise level for s values smaller than 4, so far.
Therefore, this mistake has not provided any significant error in the results
of the authors to date.

\subsection{New frequency finding methods}\label{sec_newmethod}

In common to other frequency finding methods, in our approach
the light curve data are folded with a number of trial frequencies (periods).
The trial frequency that produces the smallest scatter in the phase diagram
according to some criteria is taken as the estimate of the real period of
variations. In our approach we combine data from multiple bands with special transformations
and characterize the phase diagram scatter using variability indices calculated
from correlations of the phases (rather than correlations in the observation
times, as they were used in previous works). The even-statistic \citep[for more details see Paper II-][]{FerreiraLopes-2017papII} was used to calculate the mean, median and deviation values. It only requires that the data must have even number of measurements (and if there is an odd number, the median value is not used) while the equations to compute these parameters are equal to the previous ones. This statement will be more important when the data has only a few measurements. On the other hand, these parameters assume equal values to the previous ones when the data has even number of measurements and they are quite similar for large data samples (typically bigger than 100).

Consider a generic time series observed in multiple wavebands with observations not necessarily taken simultaneously in each band. This could also mean a time series of the same sources taken by different instruments in single or multi-wavelength observations. The sub-index $w$ is used to denote each waveband. Using this notation, all data are listed in a single table where the $i^{th}$ observation is $\mathrm{\left[ t_{i,w},\, \delta_{i,w}
\right]}$ where $\delta_{i,w}$ is given by

\begin{equation}
   \delta_{i,w} = \sqrt{\frac{n_{w}}{n_{w}-1}}\cdot \left( \frac{y_{i,w}-\bar{y_{w}}}{\sigma_{i,w}} \right),
\label{delta_stet}     
\end{equation}

\noindent  where $n_{w}$ is the number of measurements, $y_{i,w}$ are the flux
measurements, $\overline{y_{i,w}}$ is the even-mean computed using those
observations inside of a $3\sigma$ clipped absolute even-median deviation
\citep[for more details see Paper II-][]{FerreiraLopes-2017papII}, and
$\sigma_{i,w}$  denotes the flux errors of waveband $w$.  One should note that greater success in searching for signals in multi-band light curves is found when a penalty and an offset between the bands are used \citep[for more detail see][]{Long-2014,VanderPlas-2015,Mondrik-2015}. However, these modifications require a more complex model since more constraints need to be added. For our purpose, we suppose that all wavebands used are well populated in order to provide a good estimation of the mean value.

The vector given by $\mathrm{\left[ t_{i,w},\, \delta_{i,w} \right]}$ values contains data measured in either a single or multi-wavebands. For instance, w assumes a single value if a single waveband is used. Therefore the w sub-index is only useful to discriminate different wavebands and to compute the $\delta_{i,w}$.  To simplify, the $w$ subindex is suppressed in the
following steps. Therefore, the notation regarding  all observations ($N$), observed in single or multi-wavebands  by a
single or different telescopes, is given by $\mathrm{\left[ (t_{1},\, 
\delta_{1}),(t_{2},\, \delta_{2}),\cdots,(t_{N},\, \delta_{N}) \right]}$. From which, the panchromatic ($PL^{(s)}$) and flux independent ($PK^{(s)}$) period finding indices are proposed as following;

\begin{enumerate}

  \item First, consider a frequency sampling $\mathrm{F = [f_{1},f_{2},\cdots,f_{N_{freq}}]}$.
  \item Next, the phase values  $\Phi^{'} =
  [\phi^{'}_{1},\phi^{'}_{2},\cdots,\phi^{'}_{N}]$ are computed by
$\mathrm{\phi^{'}_{i} = t_{i}\times f_{1} - \lfloor t_{i}\times f_{1} \rfloor} $, where $t_{i}$ is the time and the $\mathrm{\lfloor  \rfloor}$ means the ceiling function of $\mathrm{t_{i}\times f_{1}}$.
  
  \item The phase values are re-ordered in ascending sequence of phase where 
  $\Phi = [\phi_{1},\phi_{2},\cdots,\phi_{N}]$ and $\phi_{i} \leq \phi_{j}$
  for  all $i < j$, and the $\delta_{i}$ are also
  re-ordered with their respective phases. Where for each phase value we have
  $\mathrm{\left[ (\phi_{1},\, \delta_{1}),(\phi_{2},\,
  \delta_{2}),\cdots,(\phi_{N},\, \delta_{N}) \right]}$.
   \item Next, the following parameter $Q$ of order $s$ is computed as;
         \begin{equation}
             Q^{(s)}_{i} = \Lambda^{(s)}_{i,\,\cdots\,,i+s-1} \sqrt[s]{ \left| \delta_{i} \, \cdots \, \delta_{i+s-1} \right|} 
             \label{eq_deltaq}     
         \end{equation}   
    \noindent where the $\Lambda^{(s)}$ function is given by
         \begin{equation}
            \Lambda^{(s)}_{i} = \left\{ \begin{array}{ll}
            +1 & \qquad \mbox{if \,\, $\delta_{i} > 0, \,\, \cdots \,, \, \delta_{i+s-1} > 0$ };\\
            +1 & \qquad \mbox{if \,\, $\delta_{i} < 0, \,\, \cdots \,, \, \delta_{i+s-1} < 0$ };\\
             0 & \qquad \mbox{if \,\, $\delta_{i} = 0, \,\, \cdots
             \,,\, \delta_{i+s-1} = 0$ };\\
            -1 & \qquad \mbox{otherwise}.\end{array} \right. 
            \label{eq_signal}    
         \end{equation} 

Since we are assuming that the variables are periodic, when the period is
correct, $\phi\sim0$ should be equivalent to $\phi\sim1$, so the last phases can
be correlated with the first phases, i.e. if $s=2$, $Q^{(s)}_N$ correlates
$\delta_N$ with $\delta_1$, and if $s=3$, $Q^{(s)}_{N-1}$ correlates 
$\delta_{N-1}$ with $\delta_N$ and $\delta_1$, and $Q^{(s)}_{N}$ correlates 
$\delta_{N}$ with $\delta_1$ and $\delta_2$, and so on. This consideration
ensures  the non repetition of any term and keeps the number of $Q^{(s)}$ terms 
equal to the number of observations. The sub-index $s$ sets the number of
observations that will be combined \citep[for more details see Paper I 
-][]{FerreiraLopes-2016papI}. 
     \item Finally, the period indices, equivalent to the flux-independent and panchromatic indices are given by,
    \begin{equation}
        PK^{(s)} = \frac{N^{(+)}}{N}
        \label{eq_pkind}     
     \end{equation}   

       and,
       \begin{equation}
           PL^{(s)} =  \frac{1}{N}\sum_{i=1}^{N} Q^{(s)}_{i},
           \label{eq_plind}     
       \end{equation} 
        \noindent where $N^{(+)}$ means the total number of positive correlations (see Eq.~\ref{eq_signal}). Indeed, the total number of negative correlations        ($N^{(-)}$) is given by $N^{(-)} = N - N^{(+)}$.
     \item The steps \textit{ii} to \textit{v} are repeated for all
     frequencies, $f_1$ to $f_{N_{freq}}$.
\end{enumerate}

One should be aware that $\delta_{i,w}$ values are strongly dependent on the average and hence incorrect values can be found for Algol type  variable stars and time-series which have outliers, for example. In order to more accurately measure the average value only those observations within three times the absolute even-median deviations of the even-median were used to do this. Additionally, $\Lambda^{(s)}$ is a bit different from that proposed for the flux independent variability indices. The current version assumes $\Lambda^{(s)} = 0$ if $\delta_{i} =0  \,\, \cdots \,\, \delta_{i+s-1} = 0$. This would produce $PK^{(s)}$ and $PL^{(s)}$ equal zero  in the trivial case of all observations being exactly equal, e.g. a noiseless non-variable example, i.e., $y_{i} = y_{j}$ for all $i$ values (see two last panels of Fig. \ref{fig_dist}).

\begin{figure}
  \centering
  \includegraphics[width=0.45\textwidth,height=0.3\textwidth]{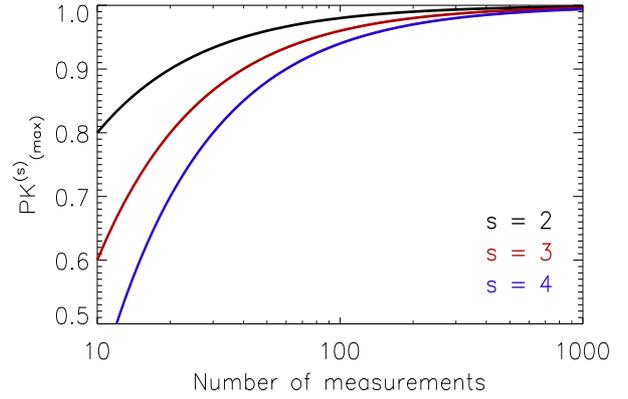} 
  \caption{The $PK_{\rm max}^{(s)}$ as function of number of measurements for a 
  sinusoidal function (see Eq. \ref{eq_pkmax}) for $s = 2$ (solid black line),
  $s = 3$ (solid red line), $s = 4$ (solid blue line) where $N^{-}_{(min)} = 2$
  was adopted.}
  \label{fig_pkmax}
\end{figure}

\subsection{The maximum $\rm PK^{(s)}$ considering different signals}\label{sec_constraint}

The maximum value allowed for the $PK^{(s)}$ parameter considering the true
variability frequency ($\mathrm{f_{true}}$) of a signal is limited by the number
of measurements which lead to $\Lambda^{(s)} = -1$, i.e., the minimum number of 
times that one of the consecutive phase observations has a value on the opposite
side of the even-mean ($N_{(c)}$). This restriction limits the maximum
value achievable by $PK^{(s)}$ ($PK_{\rm (max)}^{(s)}$). $PK_{\rm (max)}^{(s)}$ 
also varies with the order, $s$, since the number of $\Lambda^{(s)} = -1$
corresponding to observations on opposite sides of the even-mean varies with
$s$.  Indeed, $N_{(c)}$ depends on the shape of the signal.  For
instance, $N_{(c)} = 1$ for a line, $N_{(c)} = 2$ for a sinusoidal signal, 
$N_{(c)} = 4$ for a eclipsing binary light curve. Moreover, if a set of 
measurements is given by a line $y = ax+b$ ($a \ne 0 $), the number of negative
correlation measurements will be  $N^{-}_{(min)} = 1$ for $s = 2$,
$N^{-}_{(min)} = 2$ for $s = 3$, and  $N^{-}_{(min)} = 3$ for $s = 4$.
Therefore, $N^{-}_{(min)}$, and hence the maximum $PK^{(s)}$ value, varies with
s. These considerations can be expressed as following $N^{-}_{(min)} =
N_{(c)}\times\left( s - 1 \right)$. Lastly, the general relation for
$PK_{\rm (max)}^{(s)}$ can be written as

\begin{equation}
    PK_{\rm (max)}^{(s)} = 1-\frac{ N^{-}_{(min)}}{N}   = 1-\frac{ N_{(c)}\times\left( s - 1 \right)}{N}.  
   \label{eq_pkmax}     
\end{equation}

\begin{figure}
  \centering
  \includegraphics[width=0.45\textwidth,height=0.99\textwidth]{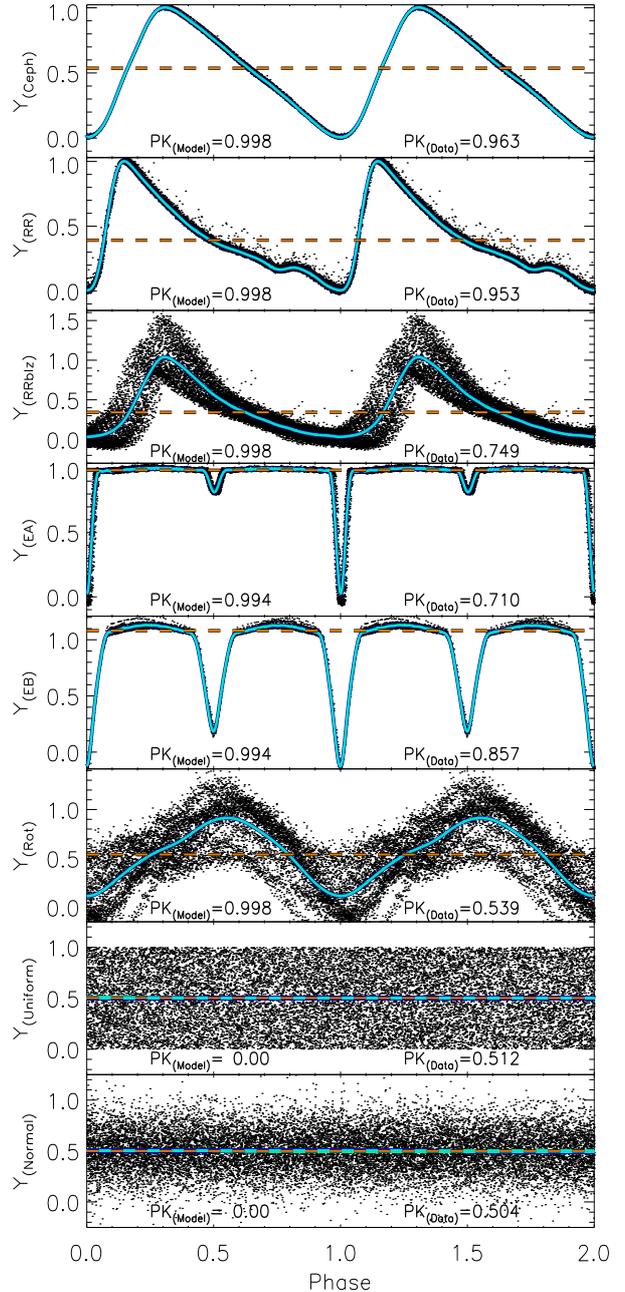} 
  \caption{Phase diagrams for pulsating stars (RR, Ceph, RRblz),
  eclipsing binaries (type EA and EB), rotational variable stars (Rot),
  and white noise from uniform and normal distributions. The data and model are
  shown as black dots and solid lines, respectively. The even-mean value
  considering those measurements within three times the absolute even-median 
  deviation is shown as orange dashed lines. Moreover, the $PK^{(2)}$ and
  $PL^{(2)}$ values for the real and modelled data are displayed at the bottom of each diagram.}
  \label{fig_dist}
\end{figure} 

A similar analytic equation for $PL^{(s)}$ index is not possible since
it depends on the amplitude. On the other hand, two features of
$PK^{(s)}$ can be seen in Eq. \ref{eq_pkmax}. First, $PK_{\rm (max)}^{(s)}$ 
values computed for two time-series having the same $N^{-}_{(\rm min)}$ value 
but a different number of observations differ (see Fig.
\ref{fig_pkmax}).
Second, all frequencies close to $\mathrm{f_{\rm true}}$ produce
$PK^{(s)} \simeq PK_{\rm (max)}^{(s)}$, since $N >> N^{-}_{(\rm min)}$.
These frequencies include the sub-harmonic frequencies of $\mathrm{f_{true}}$.  
Indeed, $\mathrm{f_{true}}$ will always return the $PK_{\rm (max)}^{(s)}$ value 
for time-series models or signals without noise (see blue lines on Fig. 
\ref{fig_dist}). However, when noise is included, statistical fluctuations can 
lead to the wrong identification of $\mathrm{f_{\rm true}}$. This means that the
$PK^{(s)}$ and consequently $PL^{(s)}$ parameters can return a main frequency 
that implies a smooth phase diagram but is different to $\mathrm{f_{\rm true}}$.

The values of $N_{(c)}$ and therefore  $N^{-}_{(min)}$ depend on the 
arrangement of observations around the even-mean and are intrinsically related 
to the signal-to-noise as discussed above (see Fig. \ref{fig_dist}).
For instance, the detached eclipsing binary (EA) signal looks like noise if only
the measurements outside of the eclipse are observed, e.g. if the phase
fraction of the eclipses is $\lesssim\frac{2}{N}$.
This means that the detection of the correct period using the $PK^{(s)}$ and
$PL^{(s)}$ parameters will be extremely dependent on the number of observations 
at the eclipses. On the other hand, the RR Lyr (RR) objects show a small 
dispersion around the even-mean even if the signal-to-noise ratio (SNR) is a 
bit low. To summarize, the $PK^{(s)}$ power will have a peak for all frequencies 
which produce smooth phase diagrams, with the largest spectrum peak
corresponding  to the smallest $N^{-}_{(min)}$. On the other hand, the $PK_{\rm
max}^{(s)}$ results in discrete values and so is more degenerate for a
small number of observations.

Figure \ref{fig_dist} shows the phase diagrams for a Cepheid (Ceph), a RR Lyrae
(RR), a RR Lyrae having the Blazhko effect (RRblz), eclipsing binaries (EA and
EB), a rotational variable (Rot), and two white noise light curves generated by 
uniform and normal distributions (for more details see Sect.
\ref{sec_simulation}). One thousand equally spaced phased measurements 
where used to plot each model and to compute the $PK^{(s)}$ shown in each panel 
($PK_{(Model)}$). For these examples, we normalize amplitude to allow us to
separate out the effects of the morphology of the light curves on these indices.
As already mentioned, $PK^{(s)}$ is not directly dependent on the
amplitude of the signal, as opposed to $PL^{(s)}$, which is.
As expected, the model crosses the even-mean twice ($N^{-}_{(min)} = 2$) for the
Ceph, RR, RRblz, and Rot models giving $PK_{\rm max}^{(2)} = 0.998$. For the 
eclipsing binaries the model crosses the even-mean four times, implying a 
$PK_{\rm max}^{(2)} = 0.996$. Actually, the EA and EB models have  $PK_{\rm max}^{(2)} = 
0.994$ due to fluctuations of the model about the even-median. Uniform and
normal distributions (i.e., time series mimicking noisy data) $PK_{\rm
max}^{(2)} = \gamma + 2^{1-s}$  where $\gamma$ is a positive number related 
with the maximum fluctuation of positive correlations. However, the $PK_{\rm
max}^{(s)} = 1$ only happens when $\delta_{i} = 0 \, \forall \, i$ values, 
i.e. noiseless non-variation.

The $PL^{(s)}$ values for the data are biased by the amplitude,
i.e., signals having different amplitudes will provide distinct values.
Consider the index values computed using the data: the EA and EB signals have
the smallest $PL^{(2)}$ values among all model tested due to their morphology,
since the majority of measurements are near to the even-mean and hence
the peak power is reduced. The Ceph, RR, and RRblz signals usually have large amplitudes and
hence large $PL^{(2)}$ values. The highest $PL^{(2)}$ values are found
for RR and Ceph signatures since there are a larger fraction of measurements
distant from the even-mean than for the other models. The $PL^{(2)}$ value for
RR stars is about half that found for the Rot model. This is a property
related to the morphology of the phase diagram.
Finally, the smallest values are found for pure-noise signals (normal
and uniform distributions).

An examination of Eq. \ref{eq_pkmax} can be used to estimate the theoretical
expected value for any signal type. However, in real data, where noise is 
included, the  $PK_{\rm max}^{(s)}$  values are smaller (see Fig.
\ref{fig_dist}) since the values decrease with the increase in the dispersion of
individual measurements about the even-mean. Therefore, $Rot$ and $EA$ models
have the largest reduction in $PK_{\rm max}^{(s)}$. In contrast, the smallest reduction of 
$PK_{\rm max}^{(s)}$ is found for Ceph and RR models since the dispersion about
even-mean is small. A detailed analysis of the weight of signal-to-noise 
ratio on $PK^{(s)}$ for different signal types is performed in the Sect. \ref{sec_simulation}.

\subsection{The optimal $s$ value}\label{sec_smax}

The optimal $s$ value ($s_{(opt)}$) will be found when
the difference between $PK^{(s)}$ computed on the phase diagram folded
using $\mathrm{f_{\rm true}}$ ($PK_{\rm max}^{(s)}$) and those ones found at
other frequencies $PK_{\rm f_{other}}^{(s)}$ is maximum. This difference can be
written as,

\begin{equation}    
    PK_{\rm (\mathrm{f_{\rm true}})}^{(s)} - PK_{\rm (f_{other})}^{(s)} \simeq PK_{\rm max}^{(s)} - PK_{\rm noise}^{(s)} 
   \label{eq_smax0}     
\end{equation}

\noindent where we consider that $PK_{\rm (f_{other})}^{(s)} \simeq
PK_{\rm noise}^{(s)}$ and compare the expressions \ref{eq_pkmax} and \ref{eq_smax0}

\begin{equation} 
  \frac{ N^{+}_{\rm f_{true}} }{ N } \simeq  1-\frac{ N_{(c)}\times\left( s_{(\rm opt)} - 1 \right)}{N}
   \label{eq_smax01}     
\end{equation}

\noindent and hence,

\begin{equation}    
   s_{(\rm opt)} \simeq  1+\frac{\left( N - N^{+}_{\rm f_{true}}  \right) }{N_{(c)}},
   \label{eq_smax1}     
\end{equation}

\noindent where $N^{+}_{\rm f_{true}}$ is the number of positive
correlations for $\mathrm{f = f_{\rm true}}$. Equation \ref{eq_smax1} provides the $s$ value 
where the maximum difference between the $PK_{\rm (\mathrm{f_{\rm
true}})}^{(s)}$ and $PK_{\rm (f_{other})}^{(s)}$ is found.

For high-SNR light curves, i.e. $N - N^{+}_{\rm f_{\rm true}}
\rightarrow N_{(\rm c)}$, $s_{(\rm opt)}=2$ since for this case $N^{+}_{\rm
f_{\rm true}} \rightarrow N$. Indeed, the $N^{+}_{\rm f_{\rm true}}$ is 
directly proportional to the SNR while $N_{(c)}$ is the opposite, i.e. the
increase of SNR increases $N^{+}_{\rm f_{true}}$ and decreases $N_{(c)}$.
Therefore, for low-SNR $N^{+}_{\rm f_{\rm true}}\rightarrow N/2$ and hence $s_{(\rm opt)} \approx 1 + 
N/2 \times N_{(\rm c)}$.  However, at the limit, $N_{(c)}$ also tends to $N/2 $
and  hence $s_{(\rm opt)} \approx 2$. To summarize, the choice of $s$ value
depends on  the signal type and SNR since $N_{(\rm c)}$ and $N^{+}_{\rm f_{\rm true}}$ 
vary with both parameters. For instance, a large value of $N_{(\rm c)}$ is 
expected for EA binary systems whatever its SNR and hence a small $s$ value is 
recommended to increase the range of signal type detected. The choice of $s$
value must take all of these properties into account.

\section{Numerical tests and simulations}\label{sec_simulation}

Artificial variable stars were simulated using a similar set of models as those 
produced in paper III \citep[for more details see][]{FerreiraLopes-2018papIII}.
Seven simulated time series were created that mimic rotational variables 
($Y_{(Rot)}$), detached eclipsing binaries ($Y_{(EA)}$), eclipsing binaries
($Y_{(EB)}$),   pulsating stars ($Y_{(Ceph)}$, $Y_{(RR)}$, $Y_{(RRblz)}$), and 
white noise ($Y_{(Uniform)}$ and $Y_{(Normal)}$). The Ceph, RR, RRblz, EA, and
Rot models were based on the CoRoT light curves \textit{CoRoT-211626074}, 
\textit{CoRoT-101370131}, \textit{CoRoT-100689962}, \textit{CoRoT-102738809}, 
and \textit{CoRoT-110843734}, respectively. The variability types were
previously  identified by \citet[][]{Debosscher-2007,Poretti-2015,Paparo-2009,
Chadid-2010,Maciel-2011,Carone-2012} and \cite[][]{DeMedeiros-2013} while the
variability period and amplitudes were reviewed by
\citet[][]{FerreiraLopes-2018papIII}. The models of variable stars were found
using harmonic fits having $12$, $12$, $12$, $24$, $24$, and $6$ 
coefficients for $Ceph$, $RR$, $RRblz$, $EA$,  $EB$, and $Rot$ variable
stars, respectively. The white noise simulations given by a normal distribution 
were used to determine the fractional fluctuation of noise (FFN). The Ceph, RR, 
RRblz, EA, and Rot models were used to realistically test and illustrate our approach.

The efficiency rate of any frequency finding method depends mainly on the signal type, the signal-to-noise ratio, and the number of observations. Therefore, three sets of simulations having $20$, $60$, and $100$ measurements for an interval of $\rm SNR$ (see Eq. \ref{eq_snr}) ranging from $\sim1$ to $\sim20$ were created for the models found in Fig. \ref{fig_dist}. In particular, $20\%$ of measurements were randomly 
selected at the eclipses for EA and EB simulations. This is required because 
these simulations look like noise if no measurement is found at the eclipses,
and is justified because any light-curves that are processed with
period-finding algorithms in NITSA must already have been selected as
variables, so eclipsing binaries with few measurements must have a relatively
high fraction at the eclipses. There will be a selection effect against binaries
with narrow eclipses, since the probability of them being detected as variables
is reduced. Values sorted randomly from a normal distribution were used to add 
noise to the simulations and the error-bars were set to be  the differences between the model and simulated data. The error bars are
not relevant to compute the $PK^{(s)}$ values. However, they are 
necessary to determine $PL^{(s)}$ parameters. The $\rm SNR$ was computed as,

\begin{equation}
   SNR = \frac{A}{2.96\times eMAD(\delta_{y})}
   \label{eq_snr}
\end{equation}

\noindent where $A$ is the signal amplitude, $\delta_{y}$ are the
residuals  (observed minus its predicted measurement), and $eMAD$ is the
even-median of the absolute deviations from the even-median. The $eMAD$ is a
slight modification of median absolute deviation from median ($MAD$).  The
$2.96\times eMAD(\delta_{y})$ is equivalent to two times the standard deviation
but it is a robust estimate of the standard deviation when outliers are 
considered \citep[e.g.][]{Hoaglin-1983}. For completeness, other estimates of
the SNR where the model is not required were tested
\citep[e.g.][]{Rimoldini-2013}. According to our tests, the latter 
usually overestimates the SNR compared with those values computed by Eq. \ref{eq_snr}.

\begin{figure}
  \centering
  \includegraphics[width=0.48\textwidth,height=0.6\textwidth]{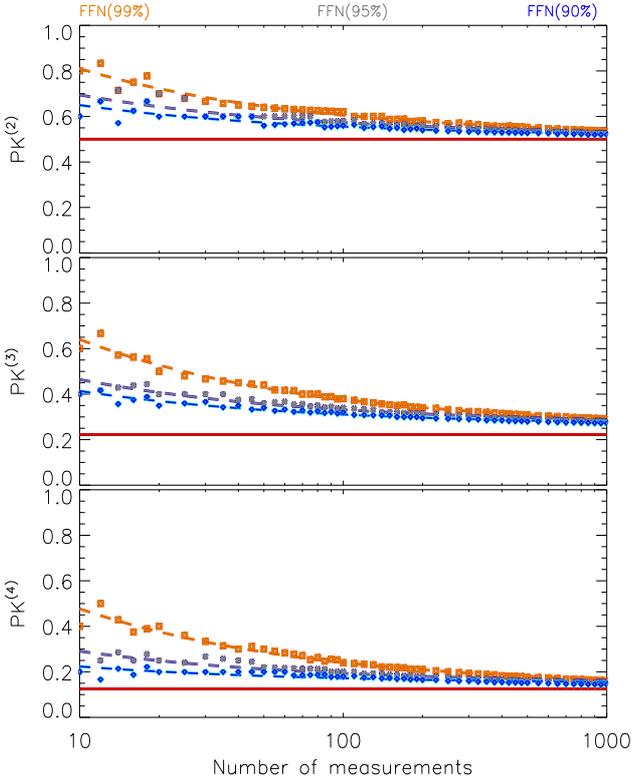}
  \caption{$PK^{(s)}$ as a function of the number of measurements for orders 2,
  3, and 4. The results for significance levels of $99\%$ (orange), $95\%$
  (grey),  and $90\%$ (blue) are in different colors. The dashed lines indicate  the $\rm FFN_{(s)}$ for models while the solid, red line shows the expected  value for the noise (see Sect.\ref{sec_fap}).}
  \label{fig_fap}
\end{figure}

\subsection{Fractional Fluctuation of Noise (FFN)}\label{sec_fap}

The fractional fluctuation of noise for signal detection is related to the level for which the figure of merit of the methods (e.g., power, in the classical periodogram) is not expected to exceed more than a fraction of times due to 
stochastic variation (or noise) on the input light curve. Indeed, the FFN 
mimics a false alarm probability since it sets the power value above which a 
certain percentage of spurious signals are found.
Indeed, there are many difficulties of estimating FAPs in realistic astronomical time series \citep[for more detail see][]{Koen-1990,Sulis-2017,VanderPlas-2018} and hence
FFN only means the lower empirical limit to find a reliable signal. 
The expected value of the flux-independent index, $K_{\rm fi}^{(s)}$, for white noise is analytical
defined as $P_{s} = 2^{s-1}$ (see Sect. \ref{sec_newmethod_intro}). The same
equation can be applied to $PK^{(s)}$ since $K_{\rm fi}^{(s)}$ and $PK^{(s)}$
are  based on the same concept. Therefore, the  $FFN_{(s)}$ can be defined as,

\begin{equation}
   FFN_{(s)} = P_{s} + \Delta = \sqrt{\frac{\alpha}{N}} + \beta
   \label{eq_fap}     
\end{equation}

\noindent  where $\alpha$ and $\beta$ are real positive numbers. $\beta$ must be
larger than  $P_{s}$ since it is a threshold for white noise. $10^{7}$ Monte
Carlo  simulations using a normal distribution were run with the number of
measurements  ranging from $10$ to $1000$ in order to compute the free
parameters for  Eq. \ref{eq_fap}. Figure \ref{fig_fap} shows the mean values of
$PK^{(s)}$ above which $1\%$ (orange dots), $5\%$ (grey dots), and $10\%$ 
(blue dots) of simulated data are found. The $FFN_{(s)}$ models are shown as dashed 
lines and the free parameters of the models are presented  in
Table \ref{tab_fap}. The minimum values of $FFN_{(s)}$ are found when  $N
\rightarrow \infty$. For this condition the $FFN_{(s)}$ estimates have values
above the noise (see Table \ref{tab_fap}). The scatter found for small numbers
of measurements (typically less than $20$) is related to the discrete values 
allowed for $PK^{(s)}$ \citep[for more details see
][]{FerreiraLopes-2016papI}. The results shown in  Fig. \ref{fig_fap} are quite similar for all uncorrelated zero-mean noise distributions.

\begin{table}
\caption[]{The constraints to $\rm FFN_{s}$ models (see Eq. \ref{eq_fap}) which delimit  $99\%$, $95\%$, and $90\%$ of white noise, respectively.}\label{tab_fap}    
 \centering 
 \begin{tabular}{| c |  c  c | c  c | c  c |}        
 \hline                
 \multicolumn{1}{|c}{  }  &  \multicolumn{2}{|c}{$99\%$}  & \multicolumn{2}{|c|}{$95\%$} & \multicolumn{2}{|c|}{$90\%$}  \\ 
 \hline               
 order  &  $\alpha$ & $\beta$ & $\alpha$ & $\beta$ & $\alpha$ & $\beta$  \\
 \hline 
$\rm  FFN_{(2)}$ & $0.9459$ & $0.5107$ & $0.5350$ & $0.5177$ & $0.4377$ & $0.5112$  \\
$\rm  FFN_{(3)}$ & $1.2321$ & $0.2541$ & $0.6150$ & $0.2696$ & $0.4784$ & $0.2625$  \\
$\rm  FFN_{(4)}$ & $1.0937$ & $0.1316$ & $0.4511$ & $0.1482$ & $0.2380$ & $0.1482$  \\
\hline 
\end{tabular}
\end{table}

The $FFN_{(s)}$  can be used as a reference to remove unreliable signals 
that lead to random phase variations in any survey, whatever the 
wavelength observed. This property is related to the weak dependence of
$PK^{(s)}$ on amplitude, error bars, or outliers according to 
\citet[][]{FerreiraLopes-2017papII}. Indeed, spurious periods that lead
to smooth phase diagrams will break this constraint. On the other hand, the
period that produces the main peak in the periodogram can be related with 
a phase diagram which has gaps for common methods like PDM and LSG.
This happens because the function used to measure the periodogram can
interpret this arrangement of measurements as a smooth phase diagram. This
result can lead to the highest periodogram peak when the signal is not well
defined and/or when a small number of epochs are available. On the other hand, 
the periods that lead to folded phase diagrams with gaps may not have many
correlated measurements and hence they will not leads to peaks in the 
$PK^{(s)}$  and $PL^{(s)}$  periodogram. Indeed, the main peak of the periodogram will be the arrangement of measurements that leads to the largest
correlation value.

\begin{figure*}
  \centering
  \includegraphics[width=0.99\textwidth,height=0.9\textwidth]{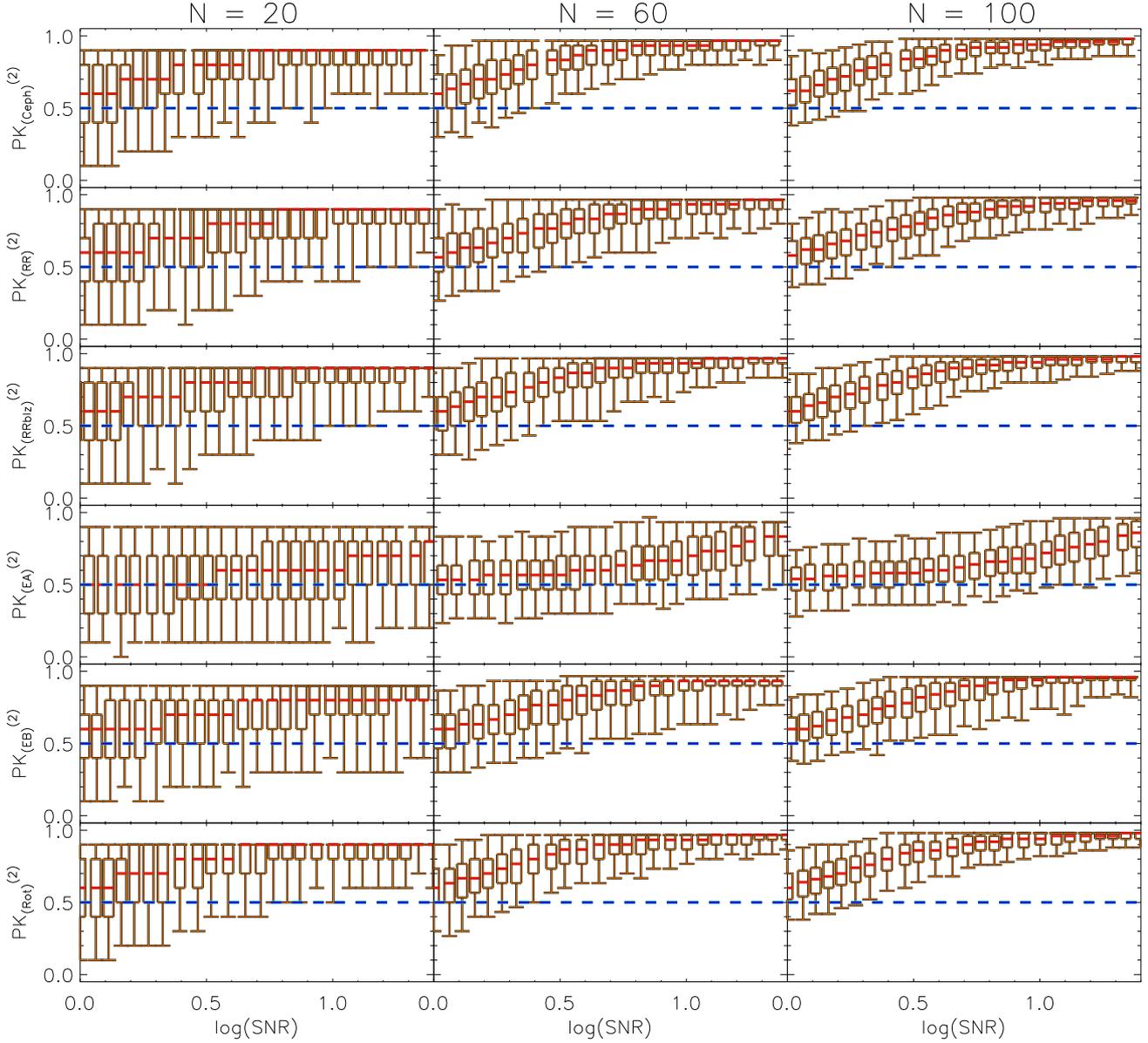}
  
\caption{$PK^{(s)}$ as a function of signal-to-noise ratio for order 2 
($s=2$) for the variable stars models shown in the Fig. \ref{fig_dist}. Each
column displays the results for $20$ (left column of panels), $60$ (middle
panels), and $100$ (right-hand columns of panels) measurements while each row
presents the results for different variable types: Cepheid, RRlyrae, RRblz, EA,
EB, and Rot models. Box plots containing $90\%$ of the data are shown. The 
even-median values for each box are marked by a solid red line while
the dashed blue lines show the expected value for the noise.}
  \label{fig_limpk}
\end{figure*}

\subsection{Dependency on the signal-to-noise ratio}\label{sec_constraintsSNR}

The $Ceph$, $RR$, $RRblz$, $EA$,  $EB$, and $Rot$ models (for more details see
Sect. \ref{sec_simulation}) were used to analyze the $PK^{(s)}$ values 
for the main variability signal. $PK^{(s)}$ values were computed using
$10^{7}$ Monte Carlo simulations for SNR ranging from $1$ to $20$. The
simulations were created for  $s = 2$, $s = 3$, and $s = 4$. The 
results for $s = 3$ and $s = 4$ show lower efficiency than those found for $s
= 2$ for lower SNR values, as expected from Sect. \ref{sec_smax}. The results
for larger orders, $s$, provide better results than those found for $s
= 2$, for high signal-to-noise time-series, having large number of measurements 
(see Sect. \ref{sec_smax}). Therefore, we only show the results for $s = 2$.
Figure \ref{fig_limpk} shows the $PK^{(s)}$ as function of SNR for $s = 2$. The 
results are displayed using box plots instead of error bars because
$PK^{(s)}$ results in discrete values and its distribution is not symmetric. A
box plot range that includes $90\%$ of results was used, and the red line
sets the middle of the distribution. The main results can be summarized as
follows:

\begin{itemize}
 \item The maximum value achieved by $PK^{(s)}$ is limited by the number of 
 measurements for all SNR. Moreover, this effect is also observed for higher s 
 orders in agreement with the values estimated by Eq. \ref{eq_pkmax} 
 (see Fig. \ref{fig_pkmax}).
  \item $PK^{(s)}$ tends to $PK_{(max)}^{(s)}$ for simulations  using $20$, 
  $60$, and $100$ measurements and high SNR for Ceph, RR, RRblz, EB, and Rot
  models.  The same trend having a slower growth is also observed for 
  EA. Indeed, $PK^{(s)}$ values are improved for EA models when the number of 
  measurements, mainly at the eclipses, is increased. About $\sim50\%$ of 
  $PK^{(s)}$ values for SNR $= 3$ are found below the expected noise
  value when the time-series has 20 measurements. This number is
  reduced to less than $\sim10\%$ when more than 60 measurements are
  available.
  \item The dispersion of $PK^{(s)}$ values decreases with the number of 
  measurements for all values of SNR. The effect is less noticeable for EA 
  models. This happens because the simulated time series looks like 
  noise when most of the measurements are sorted outside of eclipses.
  \item  About $\sim95\%$ of $PK^{(2)}$ values are above $P_{2}$ values for the
  whole range of SNR for Ceph, RR, RRblz, EB, and Rot simulations using $60$ and
  $100$ measurements. This is also true for the simulations containing 20
  measurements for $\rm SNR > 2$. On the other hand, the EA model shows 
  $PK^{(2)}$ values around the noise level for the whole range of SNR on the
  simulations containing $20$ measurements. The reason for this behaviour is the
  same as explained in the last item.
 \item The time-series like EA and EB models have the lowest $PK^{(s)}$ values 
 among all models analysed.
\end{itemize}

\begin{figure*}
  \centering
  \includegraphics[width=0.95\textwidth,height=0.40\textwidth]{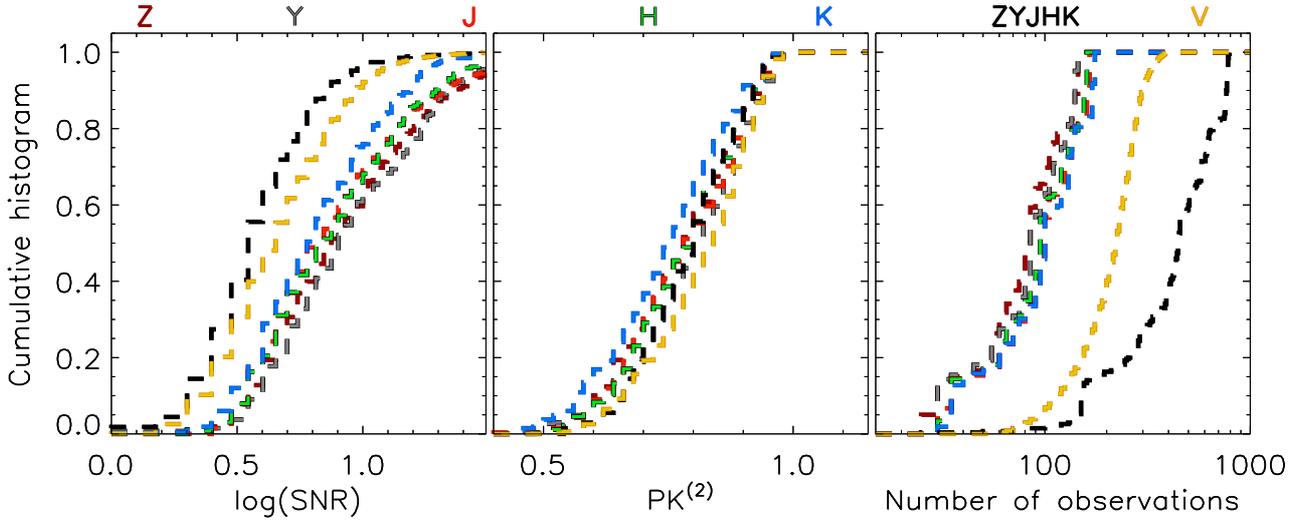}
  \caption{Cumulative histograms of SNR, $PK^{2}$, and number of measurements
  for  \textit{WVSC1} and \textit{CVSC1} stars. The results are shown for Z
  (brown),  Y (grey), J (red), H (green), K (blue), V (yellow) as well as the 
  panchromatic (black) data. }
  \label{fig_histcum}
\end{figure*}

In summary, the probability of finding $\rm PK^{(s)}$ values above the 
noise is dependent on the number of measurements, SNR, and signal type, as
expected. For all simulations, when the number of measurements is 
increased, we can measure reliable periods at lower SNR. The
simulations for higher $s$ order are quite similar to those found for  $s = 2$.

\section{Testing the method on real data}\label{sec_datatest}

A robust numerical simulation is complex because it usually does not reproduce
the correlated nature of the noise intrinsic to the data as well as variations
related to the instrumentation. Many constraints are required to provide 
realistic simulations such as a wide range of amplitudes, error bars, outliers, 
and correlated noise, to name a few. However, the simulation of the $PK^{(s)}$ 
power is facilitated because: (a) the amplitude can be a free parameter since 
$PK^{(s)}$ is only weakly dependent on it; (b) the even mean values 
(see Eq. \ref{delta_stet}) are computed using those observations within three 
times the absolute even-median deviation, which effectively reduces the outliers
weight on zero point estimation ($\overline{y_w}$, see
Eqn.~\ref{delta_stet}) but all epochs are considered to compute the powers; (c)
the correlated nature of successive measurements is reduced since they are
computed using phase diagrams. On the other hand, a robust simulation for
$PL^{(s)}$ covering all important aspects of it is difficult because 
$PL^{(s)}$ has a strong dependence on amplitude, outliers, and error
bars. Therefore, the discussions in the previous sections only address the 
constraints on $PK^{(s)}$.

The $PK^{(s)}$ and $PL^{(s)}$ methods can be tested on real data using existing
variable stars catalogues. The WFCAMCAL variable stars catalogue
(\textit{WVSC1}) having  $280$ stars \citep[][]{FerreiraLopes-2015wfcam} and the
Catalina Survey Periodic Variable  star  catalogue (\textit{CVSC1}) having
$\sim47000$ sources \citep[][]{Drake-2014} were used to estimate the efficiency 
rate of our new period finding methods. The \textit{WVSC1} was created from the
analysis of the WFCAM Calibration 08B release \citep[WFCAMCAL08B -
][]{Hodgkin-2009, Cross-2009}. More information about the design, the 
data reduction, the layout, and about variability analysis of this database are
described in detail in
\citealt[][]{Hambly-2008,Cross-2009,FerreiraLopes-2015wfcam}. The
WFCAMCAL database is a useful dataset to test single and panchromatic
wavelength period finding methods. To summarize, WFCAM database contains 
panchromatic data ($ZYJHK$ wavebands) that were observed to calibrate the UKIDSS
surveys \citealt[][]{Lawrence-2007}. A sequence of filters $JHK$ or $ZYJHK$ were
observed within a few minutes during each visit the fields. These sequences were
repeated in a semi-regular way that leads to an uneven sampling having large
seasonal gaps. On the other hand, the \textit{CVSC1} has a huge amount of
objects, with seventeen variable stars types that were visually inspected by the
authors.

In order to perform a straightforward comparison between the results, the SNR 
of \textit{WVSC1} and \textit{CVSC1} stars were also estimated using  Eq.
\ref{eq_snr}. Also, the number of measurements and the $PK^{(2)}$ values were
computed. Figure \ref{fig_histcum} shows the cumulative histograms of SNR and 
number of measurements for \textit{WVSC1} and \textit{CVSC1} stars. The results 
for each waveband as well as for panchromatic data are shown by different
colours. About $\sim90\%$ of \textit{WVSC1} single waveband has $SNR > \sim3$
while this number decreases to $\sim70\%$ for the \textit{CVSC1} and
panchromatic data. The \textit{WVSC1} single waveband data have a number of
measurements ranging from $\sim30$ to $\sim150$ while  \textit{CVSC1} stars 
have a number from $\sim100$ to $\sim300$. When the panchromatic data is 
considered, the number of measurements increases considerably by a factor  $\sim
5$ compared with \textit{WVSC1} single waveband. However, the SNR are smaller 
than those found for single wavebands. In general, the SNR for panchromatic 
wavebands are smaller than those found for \textit{CVSC1} stars.

\begin{figure*}
  \centering
  \includegraphics[width=0.49\textwidth,height=0.3\textwidth]{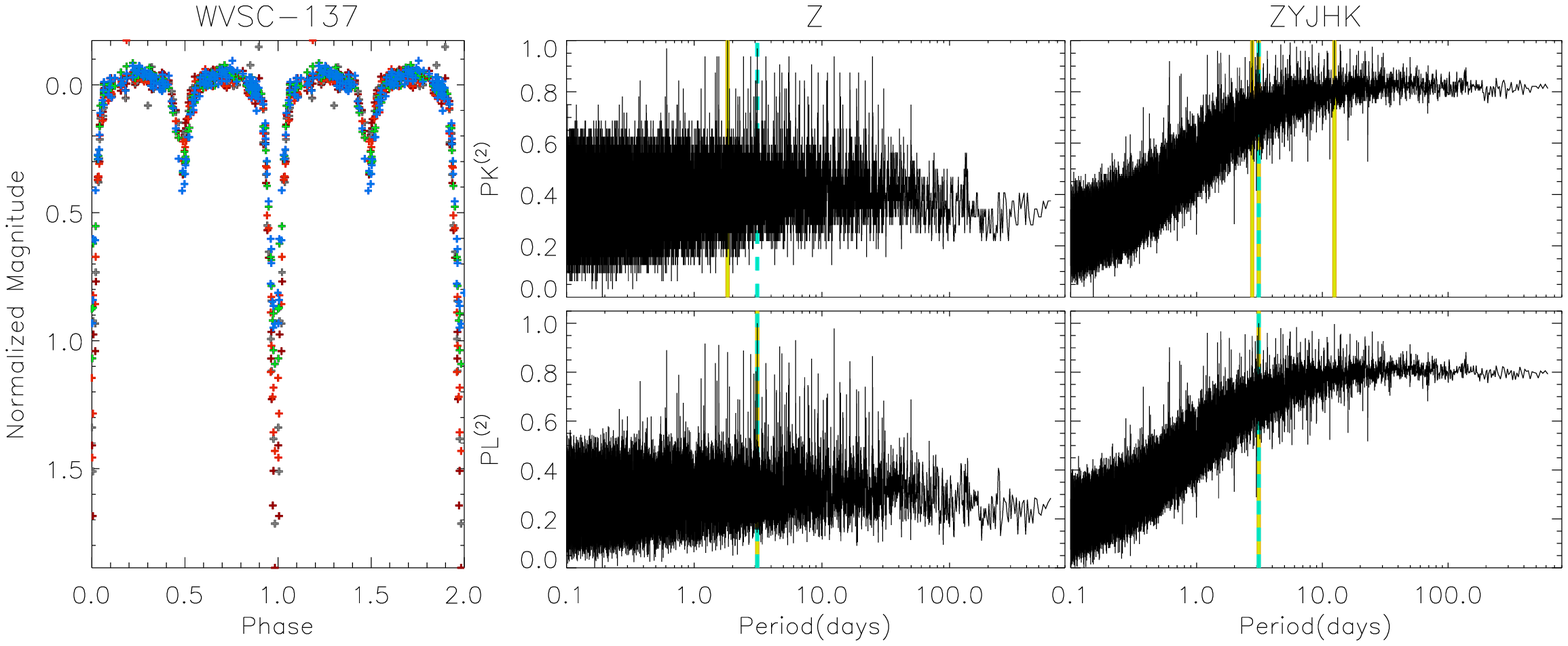}
  \includegraphics[width=0.49\textwidth,height=0.3\textwidth]{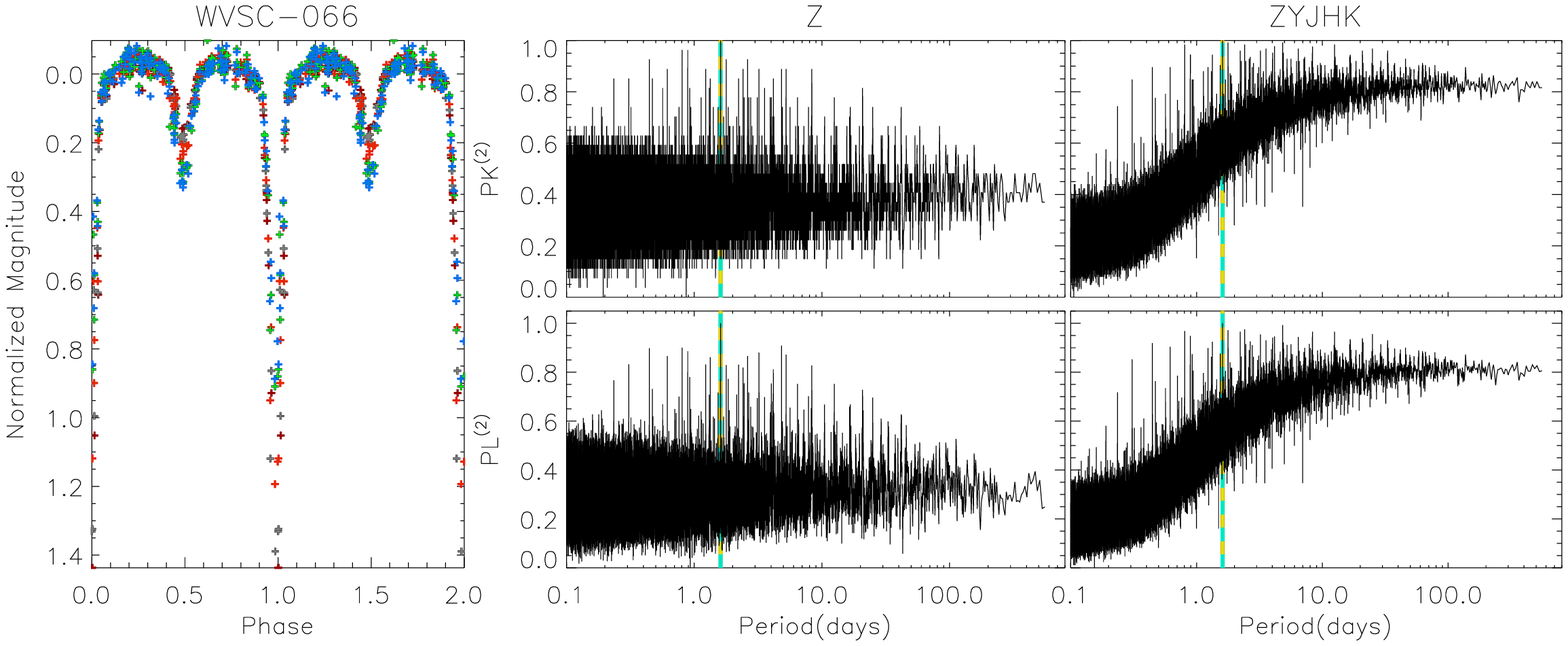}

  \includegraphics[width=0.49\textwidth,height=0.3\textwidth]{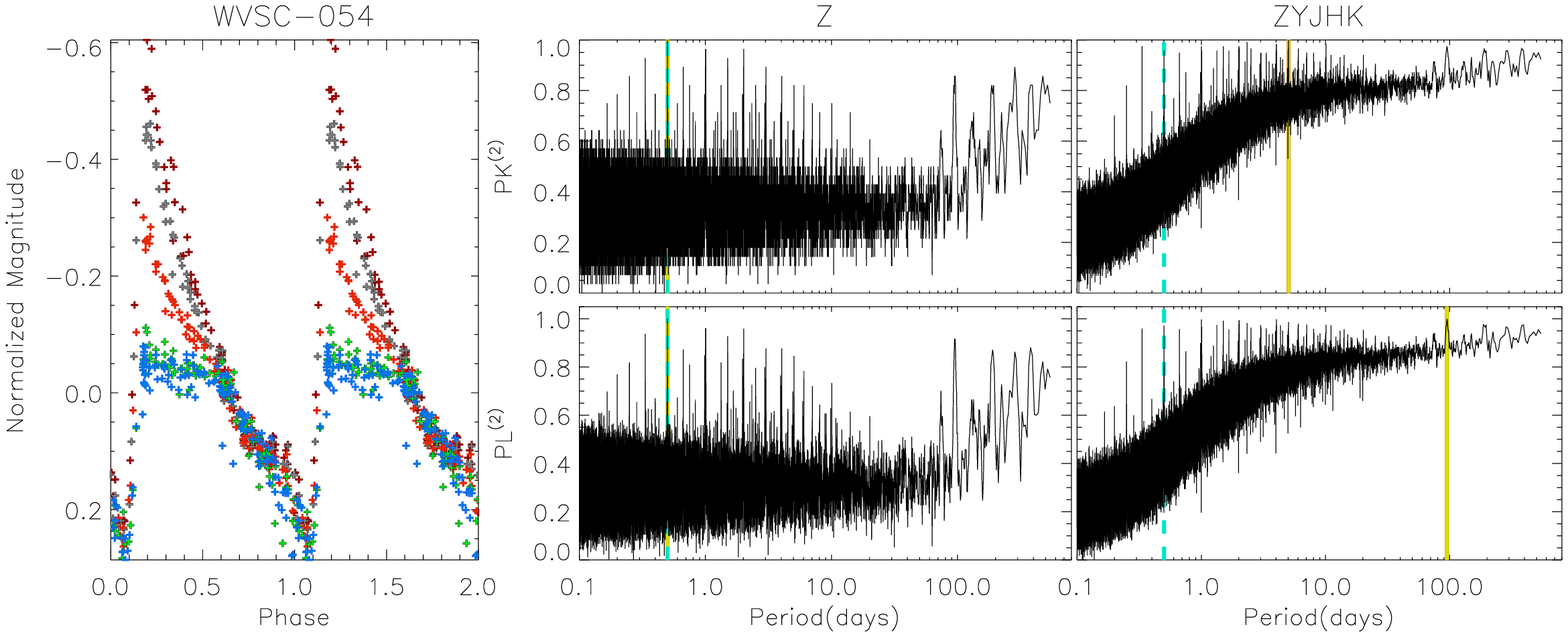}
  \includegraphics[width=0.49\textwidth,height=0.3\textwidth]{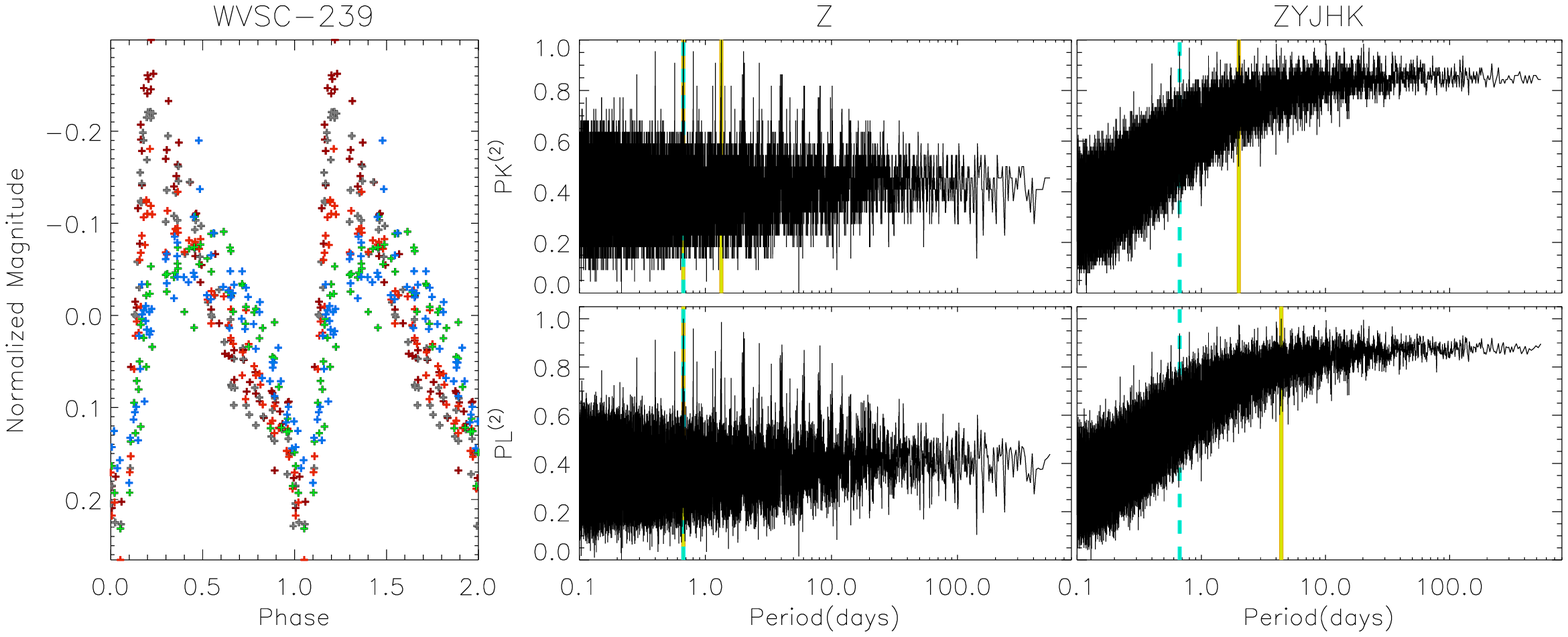}
  
  \includegraphics[width=0.49\textwidth,height=0.3\textwidth]{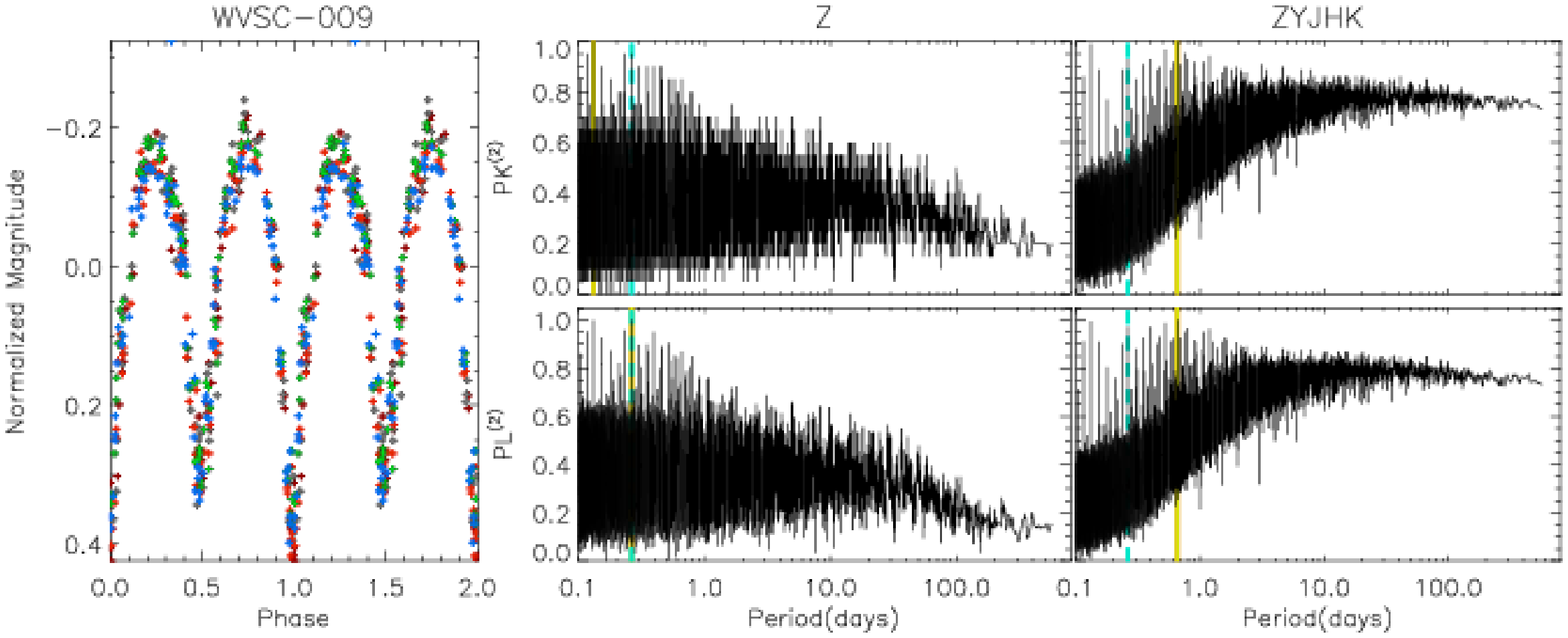}
   \includegraphics[width=0.49\textwidth,height=0.3\textwidth]{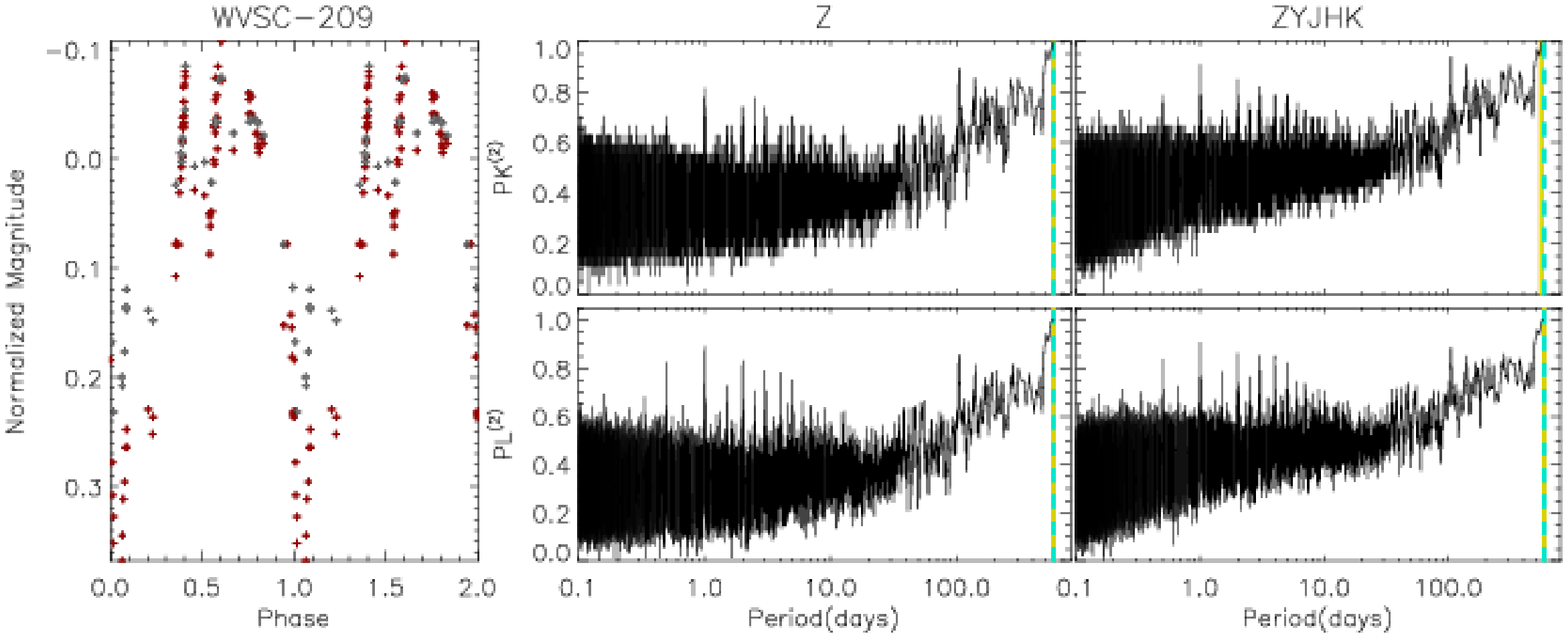}
  
  \includegraphics[width=0.49\textwidth,height=0.3\textwidth]{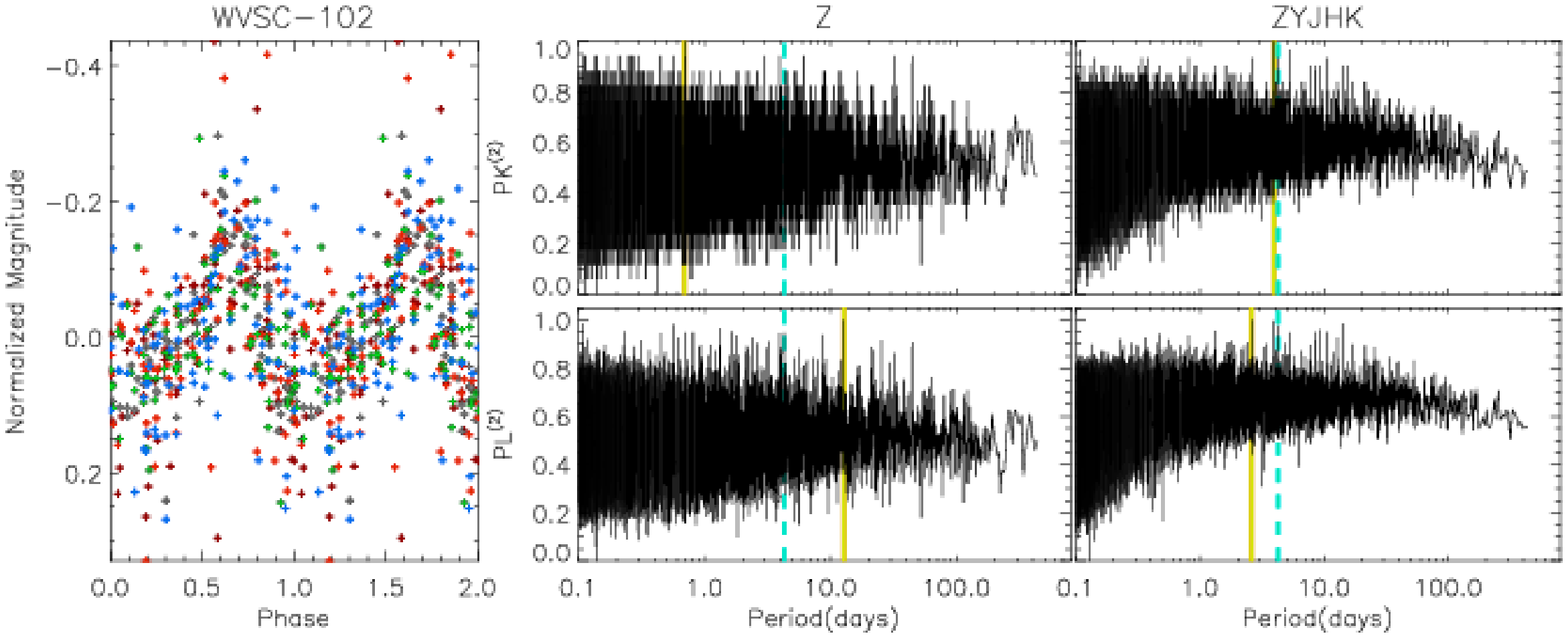}
  \includegraphics[width=0.49\textwidth,height=0.3\textwidth]{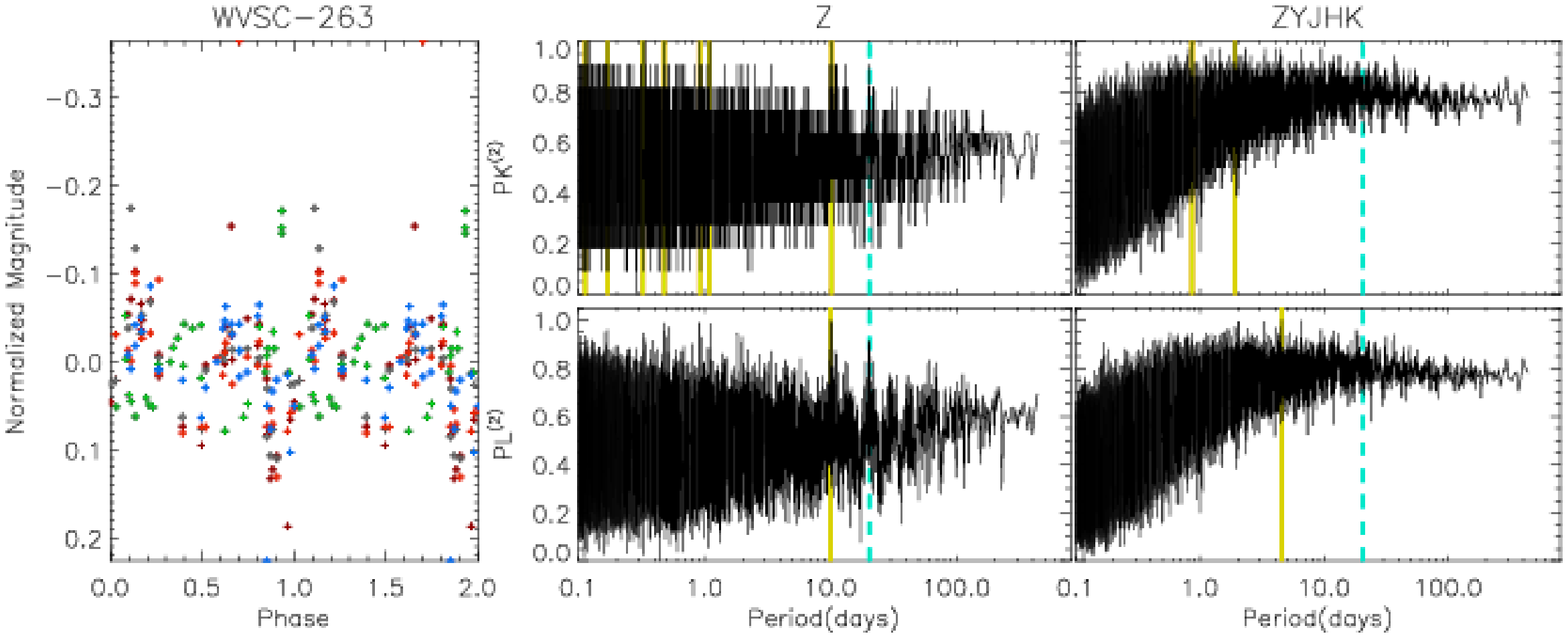}
  
  \caption{Phase diagrams of panchromatic data (left panels) and the $PK^{(s)}$
  and  $PL^{(s)}$ normalized periodogram for Z and ZYJHK wavebands
  (right panels). The cross symbols in the phase diagrams set the measurement of
  Z (brown), Y (grey), J (green), H (red), and K (blue) wavebands. The dashed
  green lines indicate the published variability periods while the full yellow
  lines  indicate the periods related with the largest peak in the periodogram.}
  \label{fig_lcpower}
\end{figure*}

Understanding the peculiarities of the sample tested is crucial when
analysing the efficiency rate of our approach. Therefore we summarize 
how the period searches were performed to find periods for \textit{WVSC1} and 
\textit{CVSC1} stars. \citet[][]{FerreiraLopes-2015wfcam} selected about $6651$
targets to which four period finding methods were applied. Next, the ten best
ranked periods in each of the four methods were selected. For each period a
light curve model was created using harmonics fits. Finally, the
very best period was chosen as that with the smallest $\chi^2$ with
respect to all ranked periods; On the other hand, the period search for $\sim
154$ thousand \textit{CVSC1} sources was made using the Lomb
Scargle method. Next, the main periods were analysed using the Adaptive Fourier 
Decomposition (AFD) method \citep[][]{Torrealba-2015} in order to determine the 
main variability period and reduce the number of sources to be visually 
inspected ($112$ thousand). Additionally, the periods of a large number of the 
sources were improved and corrected by the authors. Many of the variability 
periods  of \textit{WVSC1} and \textit{CVSC1} were related to 
sub-harmonics of their true period and the final results were set
after visual inspection.

The following sections discuss the \textit{WVSC1} and \textit{CVSC1} variable
stars from the viewpoint of $PK^{(s)}$ and $PL^{(s)}$ parameters. The periodogram, the efficiency rate, and the peculiarities of our approach are 
analysed. For that, we perform the period search using the SLM, LSG, and PDM
methods besides the panchromatic and flux independent methods. A frequency 
range of $(2/T_{max}) d^{-1}$ to $30 d^{-1}$  was explored and we evenly sampled
this frequency range with a frequency step of $\frac{1}{300\times
T_{max}}$, where $T_{max}$ is the total time span. The
frequency sampling constrains a maximum phase shift of $0.1$ that allows us
to detect the large majority of signal types (for more detail see Paper III).  A
quick visual inspection was performed on some of  \textit{WVSC1} and
\textit{CVSC1} to test our analysis in the next sections. The main goal of this
work is to propose a new period finding method instead of checking the 
reliability of the periods in the  \textit{WVSC1} and \textit{CVSC1} catalogues.

\subsection{Periodogram and efficiency rate}\label{sec_powspec}

The $PK^{(2)}$ and $PL^{(2)}$ periodogram were computed for the 
\textit{WVSC1} and \textit{CVSC1} stars. For better visualization, the
differential panchromatic light curve (see Fig. \ref{fig_lcpower}) was obtained
by subtracting the even-median from the magnitudes in each light curve. As a
result, light curves with zero mean are produced. The $PK^{(s)}$ and $PL^{(s)}$ 
parameters do not use any kind of transformation to combine measurements at
different wavelengths. However, a better way to combine multi-wavelength data is an open question.

Figure \ref{fig_lcpower} shows the phase diagrams and their normalized 
periodogram for some \textit{WVSC1} stars. The phase diagrams (left panels)
show the folded panchromatic data while the periodogram considering single 
and panchromatic wavebands are displayed in the centre and right panels
respectively. The periodogram for a large part of \textit{CVSC1} stars are
quite similar to those found for panchromatic data, i.e., periodogram for 
sources having more measurements and smaller SNR than for the \textit{WVSC1} 
single waveband. The main results can be summarized as follows:

\begin{itemize}
   \item  $PK^{(2)}$ has more than one peak related with the maximum power for
   \textit{WVSC-239} and \textit{WVSC-263}, i.e., this means that
   $PK^{(s)}_{(max)}$ indicates more than one viable period. The number of such 
   periods gives the ''degeneracy'' of a particular arrangement of measurements 
   in the phase diagram. This number increases as the number of measurements 
   decreases (see Eq. \ref{eq_pkmax}).
   
   \item The main period found by $PK^{(2)}$ is not the same as that obtained 
   by $PL^{(2)}$. This means that the maximum number of positive correlations
   is not the same as the maximum correlation power. Besides, the ''degeneracy''
   of periods found for $PK^{(2)}$ is not observed in the $PL^{(2)}$ periodogram.

   \item  $PK^{(2)}$ and  $PL^{(2)}$ periodogram for the Z waveband show a scatter around the expected noise level.   Additionally, the \textit{WVSC-054} and \textit{WVSC-209} periodogram  show an increase in $PK^{(2)}$ for long periods. Indeed, this behaviour is
   observed in all wavebands and hence it is an attribute of the proposed method.

   \item $PK^{(2)}$ for panchromatic data increases with periods up to a 
   maximum of around $5$ days and then levels off and drops slightly for longer
   periods for almost all sources. These sources have variability periods of
   less than $5$ days. This behaviour is not found for \textit{WVSC-209} and
   \textit{WVSC-102}. Indeed, \textit{WVSC-102} has a variability period of
   $589$ days and hence the trend observed is different to the others. On
   the other hand, \textit{WVSC-209} has a low-SNR signal. Therefore, this trend
   is related with the variability period and SNR. Indeed, the phase diagram
   keeps part of  the correlation information when the light curve is folded 
   using a test period bigger than the true variability period.

   \item $PK^{(2)}$ and $PL^{(2)}$ periodogram have peaks at the
   previous measured (true) variability periods of \textit{WVSC1} stars.
   However, the period related with the largest peak is not always the
   true variability period.

   \item The panchromatic data have lower SNR. Indeed, no clear signal can be
   observed in \textit{WVSC-263}. This could mean that the signal shape is very
   different from one band to another, or a signal or seasonal variation is 
   present in a single waveband, or the variability period is wrong, to name the
   most likely possibilities.

\end{itemize}

To summarize, the $PK^{(s)}$ and $PL^{(s)}$ periodogram indicate the 
arrangement of measurements in the phase diagram that maximize the correlation 
signal and power, respectively.  Therefore, the $PK^{(s)}$ and $PL^{(s)}$
parameters can be used to identify the periods that lead to a smooth phase 
diagram from the viewpoint of correlation strength.

\begin{table*}
\caption[]{Accuracy considering two approaches; the main period ($E_{(M)}$)
and the main period plus its sub-harmonic and overtone. We consider that there
is agreement if the relative difference is smaller than $1\%$.}\label{tab_erate}
 \centering 
 \begin{tabular}{| c |  c  c | c  c | c  c | c  c | c  c | c  c | c  c |}        
 \hline                
 \multicolumn{1}{|c}{  }  &  \multicolumn{2}{|c}{Z}  & \multicolumn{2}{|c|}{Y} & \multicolumn{2}{|c|}{J} & \multicolumn{2}{|c|}{H} & \multicolumn{2}{|c|}{K} & \multicolumn{2}{|c|}{ZYJHK} & \multicolumn{2}{|c|}{V}  \\ 
 \hline               
   &   $E_{(M)}$ & $E_{(MH)}$ &   $E_{(M)}$ & $E_{(MH)}$ &   $E_{(M)}$ & $E_{(MH)}$ &   $E_{(M)}$ & $E_{(MH)}$ &   $E_{(M)}$ & $E_{(MH)}$ &   $E_{(M)}$ & $E_{(MH)}$  &   $E_{(M)}$ & $E_{(MH)}$     \\
\hline 
$\rm PK^{(2)}$ & $0.30$ & $0.55$ & $0.30$ & $0.59$ & $0.30$ & $0.60$ & $0.29$ & $0.59$ & $0.23$ & $0.50$ & $0.19$ & $0.40$ & $0.24$ & $0.51$  \\
$\rm PL^{(2)}$ & $0.20$ & $0.46$ & $0.22$ & $0.49$ & $0.22$ & $0.50$ & $0.28$ & $0.57$ & $0.17$ & $0.44$ & $0.14$ & $0.27$ & $0.25$ & $0.53$  \\
LSG & $0.10$ & $0.27$ & $0.09$ & $0.29$ & $0.09$ & $0.34$ & $0.09$ & $0.26$ & $0.11$ & $0.31$ & $0.09$ & $0.32$ & $0.20$ & $0.87$  \\
PDM & $0.17$ & $0.50$ & $0.26$ & $0.59$ & $0.25$ & $0.62$ & $0.27$ & $0.62$ & $0.22$ & $0.57$ & $0.26$ & $0.69$ & $0.22$ & $0.88$  \\
SLM & $0.30$ & $0.55$ & $0.30$ & $0.59$ & $0.29$ & $0.63$ & $0.28$ & $0.59$ & $0.22$ & $0.50$ & $0.15$ & $0.32$ & $0.25$ & $0.49$  \\

\hline 
\end{tabular}
\end{table*}

\subsection{Accuracy}\label{sec_erate}

The accuracy was measured considering the main signal(s) detected by the
$PK^{(2)}$ and $PL^{(2)}$ methods. Indeed, the largest power of  $PK^{(s)}$ 
periodogram can be related to more than one period. Therefore, all periods 
related to the largest periodogram peak were considered to measure the accuracy, i.e., the recovery fraction of
variability periods. Two parameters to measure the accuracy were considered: $E_{(M)}$ - when the  main
period is detected; $E_{(MH)}$ - when the main variability period ($P_{Lit}$),
measured in \cite{FerreiraLopes-2015wfcam}, or its sub-harmonic, or overtone is found.
Indeed, the processing time of each method was not taken into account in this discussion. A new approach to reduce
the running time necessary to perform period searches will be addressed in
a forthcoming paper in this series. Those signals found within $\pm1\%$
of the variability period were considered as detected. Table \ref{tab_erate}
shows the results for individual wavebands as well as for the panchromatic data.
The main results can be summarized as:

\begin{itemize}
   \item The accuracy is lower than $100\%$ for all methods and data tested.
   However, new estimates of Catalina variability periods have been
   produced recently \citep[e.g.][]{Papageorgiou-2018} and the
   \textit{CVSC1} combines the results found by PDM, LSG, and STR methods for
   all wavebands to determine the best variability period. Therefore, the 
   accuracy for both datasets is larger than that displayed
   in Table \ref{tab_erate} if these results are taken into
   consideration. 
   
   \item $PK^{(2)}$ has the highest efficiency rate considering only the  main
   period ($E_{(M)}$) for the Z, Y, J, H, and K wavebands. The efficiency rate
   of $PK^{(2)}$ for the V waveband and panchromatic data is similar to
   that found for the SLM and $PL^{(2)}$ methods. $E_{(M)}$ decreases
   form Z to K wavebands because the first ones have larger SNR.

   \item The $E_{(MH)}$ values for Z, Y, J, H, and K for $PK^{(2)}$ and SLM are
   quite similar  and they have the highest 
 accuracy for the Z and Y wavebands. On the other hand, PDM has the highest
   $E_{(MH)}$ values for H,  ZYJHK, and V wavebands. Indeed, the $E_{(MH)}$ 
   values for PDM and LSG are quite similar for the V waveband.
         
   \item  The $E_{(M)}$ for $PL^{(2)}$ is always smaller than that found for
   $PK^{(2)}$ except for V band where it is $4\%$ lower for $PK^{(2)}$.

   \item The highest $E_{(M)}$ is found for $PK^{(2)}$ method while the highest 
   $E_{(MH)}$ is found for the PDM method. The accuracy found for LSG method
   is quite similar to that found for PDM method for V waveband while in other
   wavebands the PDM method has twice the accuracy. Indeed, this difference is
   reduced by a few percent if a higher relative error is considered. Indeed,
   the $P_{Lit}$ found by the ZYJHK wavebands are
   refined using the SLM method \citep[for more details
   see][]{FerreiraLopes-2015wfcam}. On the other hand, the V waveband results
   were computed using Lomb-Scargle and refined using reduced $\chi^2$
   \citep[for more details see][]{Drake-2014}. Therefore, the accuracy can be
   biased by the approach used to improve the variability period estimation.
   Indeed, a deep discussion about how to determine accurately the variability
   period and its error is found in the third paper of this series \citep[for
   more details see][]{FerreiraLopes-2018papIII}.

   \item The panchromatic data do not significantly increase the efficiency rate
   for any method. The panchromatic data provides a larger number of
   measurements but a smaller SNR compared with those found for single wavebands
   (see Fig. \ref{fig_histcum}).
   
   \item The efficiency rate of $PK^{(2)}$, $PL^{(2)}$, and SLM is strongly 
   decreased for V and panchromatic data. This is related with the smaller SNR 
   of these data. It indicates a strong dependence of the $PK^{(2)}$ and
   $PL^{(2)}$ methods on the SNR.
      
\end{itemize}

The periods detected by PDM are also detected by the LSG method. 
Moreover, almost all periods detected using the PK, PL, and STR methods are also
found by the LSG method. The periods detected using LSG or PDM which are not
found by other methods belong mainly to a few types: W UMa ($\sim61\%$), EA
($\sim13\%$), RR Lyr on first overtone ($\sim10\%$), and RR Lyr on several
modes ($\sim2\%$), measured using the ratio of the number of missed sources to
the total number of sources missed. Indeed, the largest miss-rate is found
for the multi-periodic RR Lyr-type when the relative number of sources are
considered, i.e. the fraction of missed sources divided by the number of sources
detected for each variability type. As expected, the multi-periodic periods have 
the largest miss-rate since the current approach is not designed to select these
periods. From quick visual inspection on phased data of the periods found by
methods other than PDM and LSG the following concerns have been raised: the
period found is a higher harmonic or overtone of that found in the literature; 
the phase diagram is not always smooth; the period found sometimes produces
a smooth phase diagram but the period found is different or has a relative
difference larger than $1\%$. This means that STR, PK, and PL methods are more
likely to return spurious periods or higher harmonics of the main variability period since
the majority of the sources missed belong to these groups.

\begin{figure*}
  \centering
  \includegraphics[width=0.32\textwidth,height=0.25\textwidth]{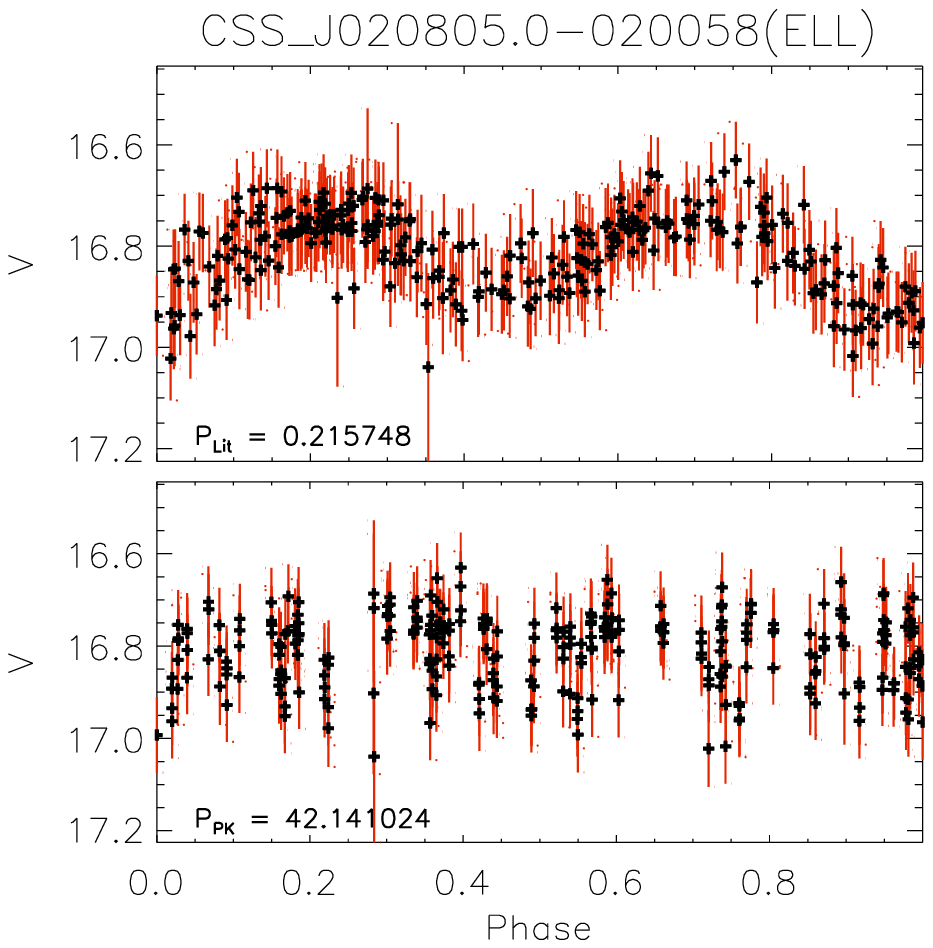}
  \includegraphics[width=0.32\textwidth,height=0.25\textwidth]{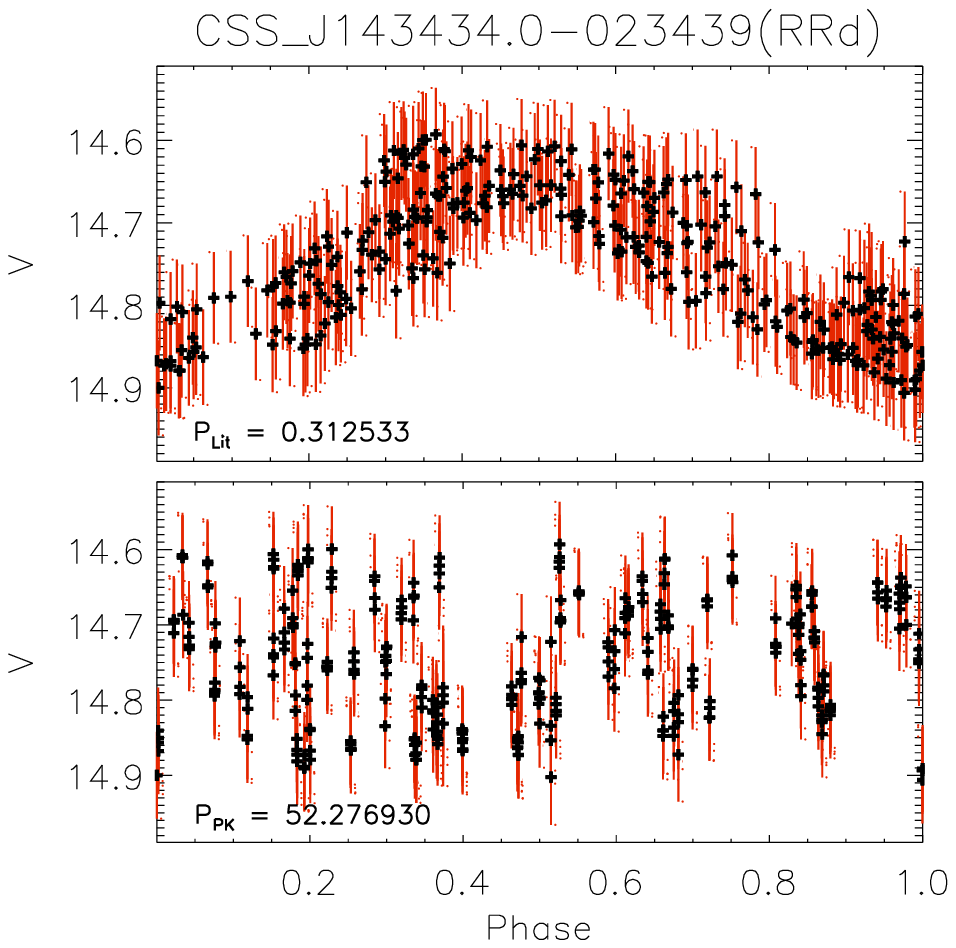}
  \includegraphics[width=0.32\textwidth,height=0.25\textwidth]{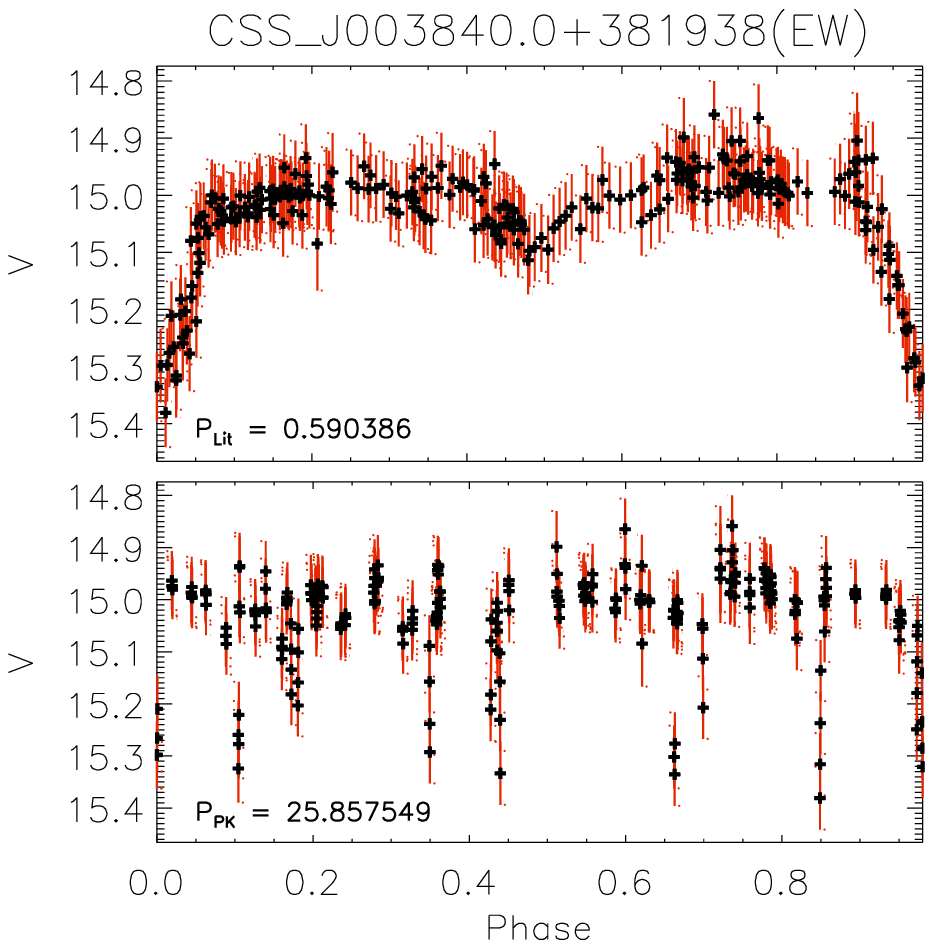}

  \includegraphics[width=0.32\textwidth,height=0.25\textwidth]{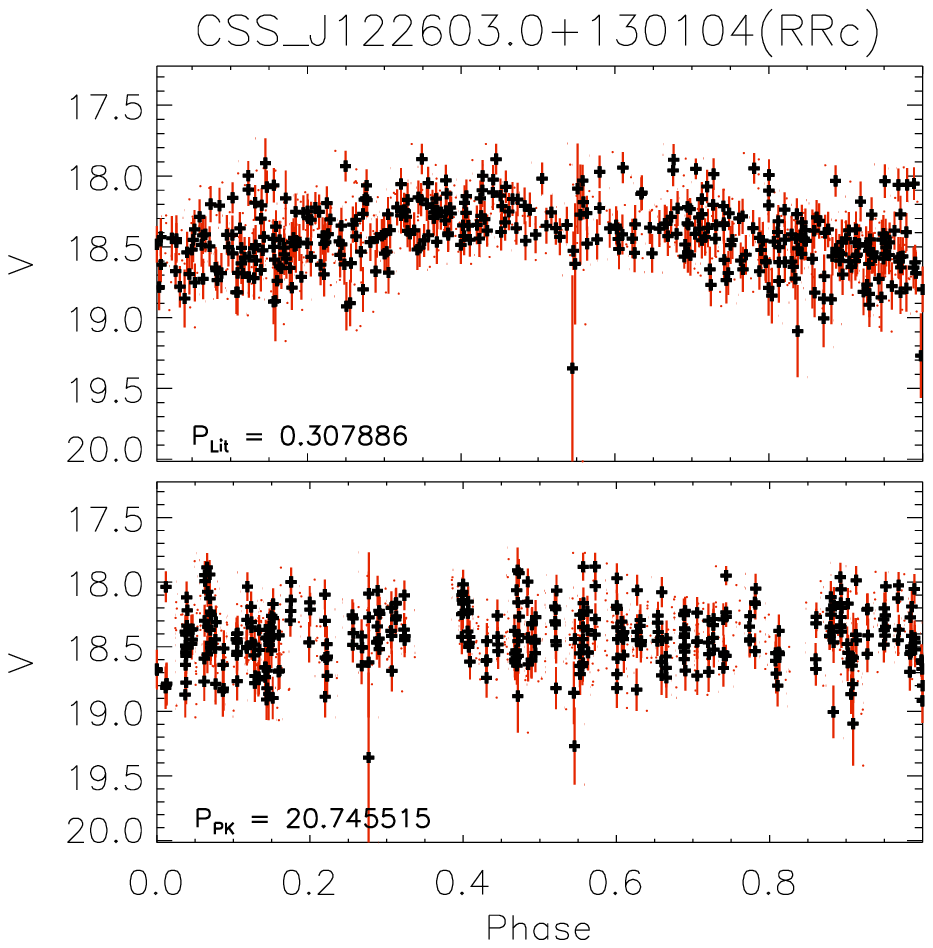}
  \includegraphics[width=0.32\textwidth,height=0.25\textwidth]{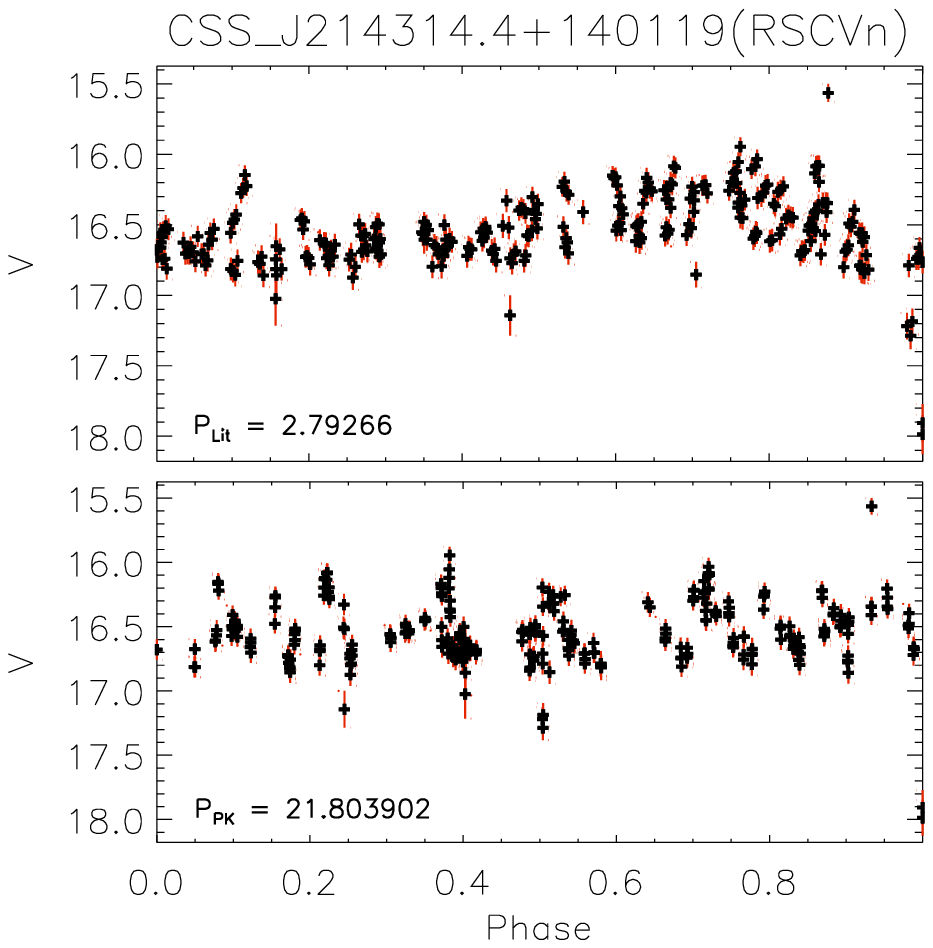}
  \includegraphics[width=0.32\textwidth,height=0.25\textwidth]{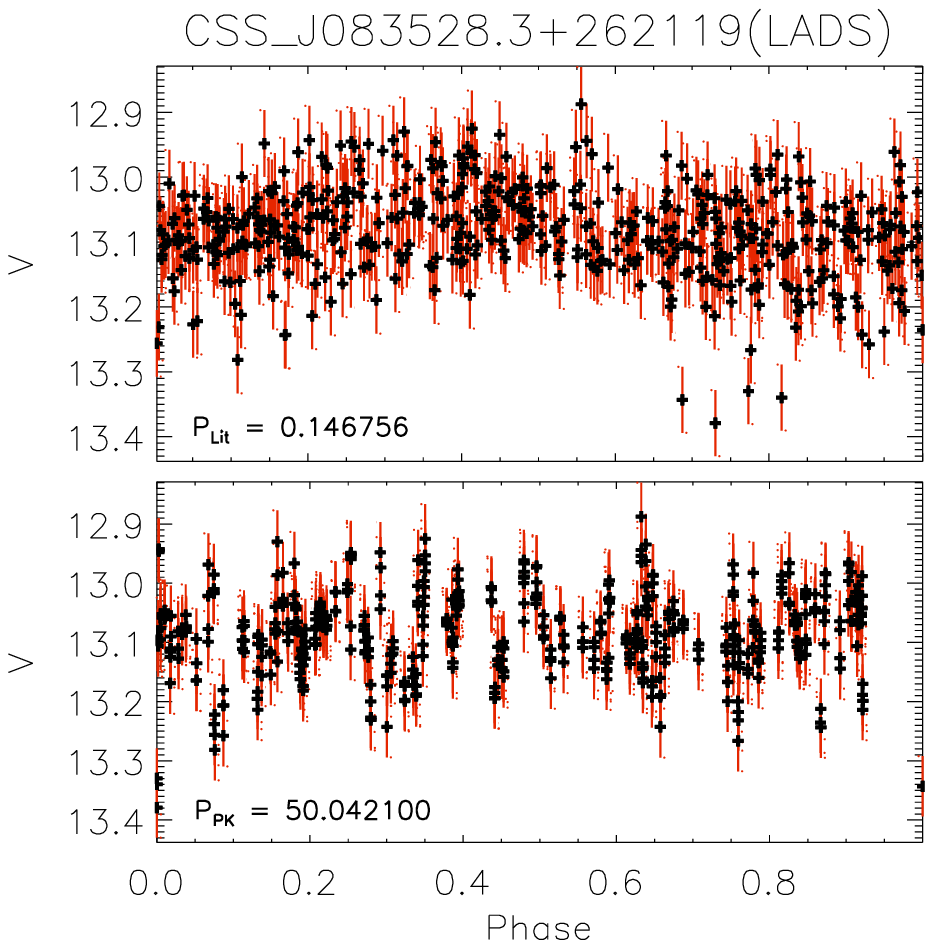}

  \includegraphics[width=0.32\textwidth,height=0.25\textwidth]{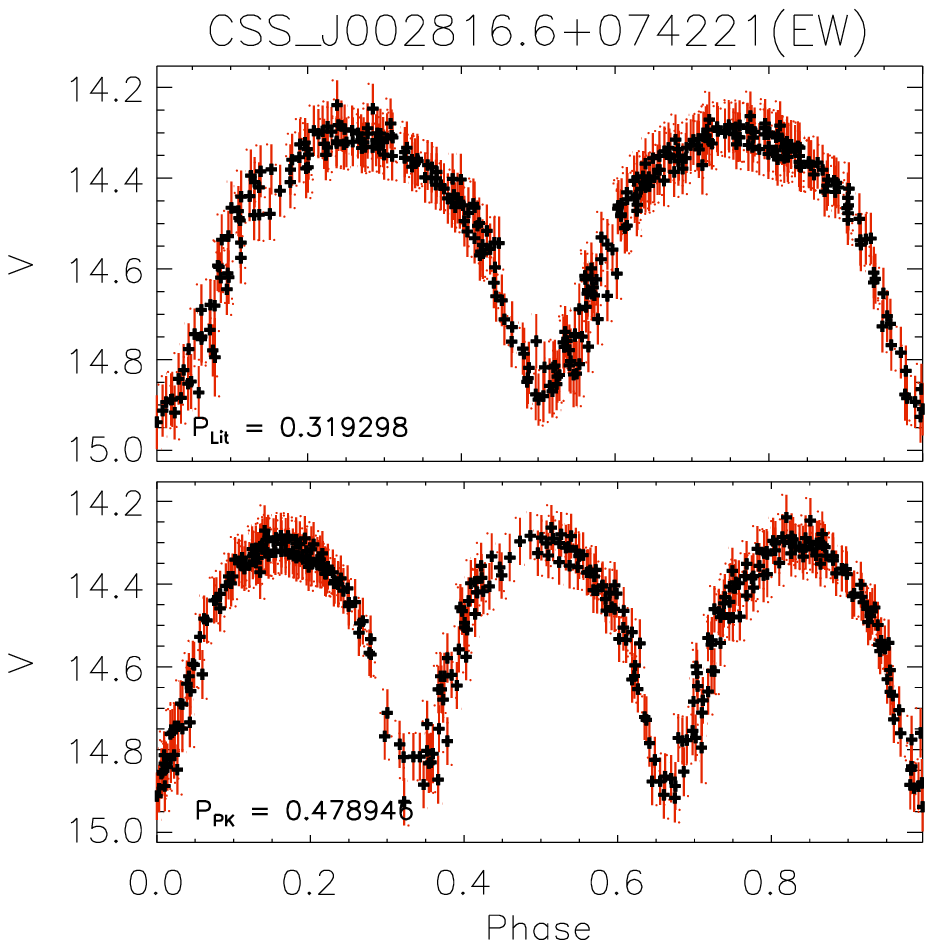}
  \includegraphics[width=0.32\textwidth,height=0.25\textwidth]{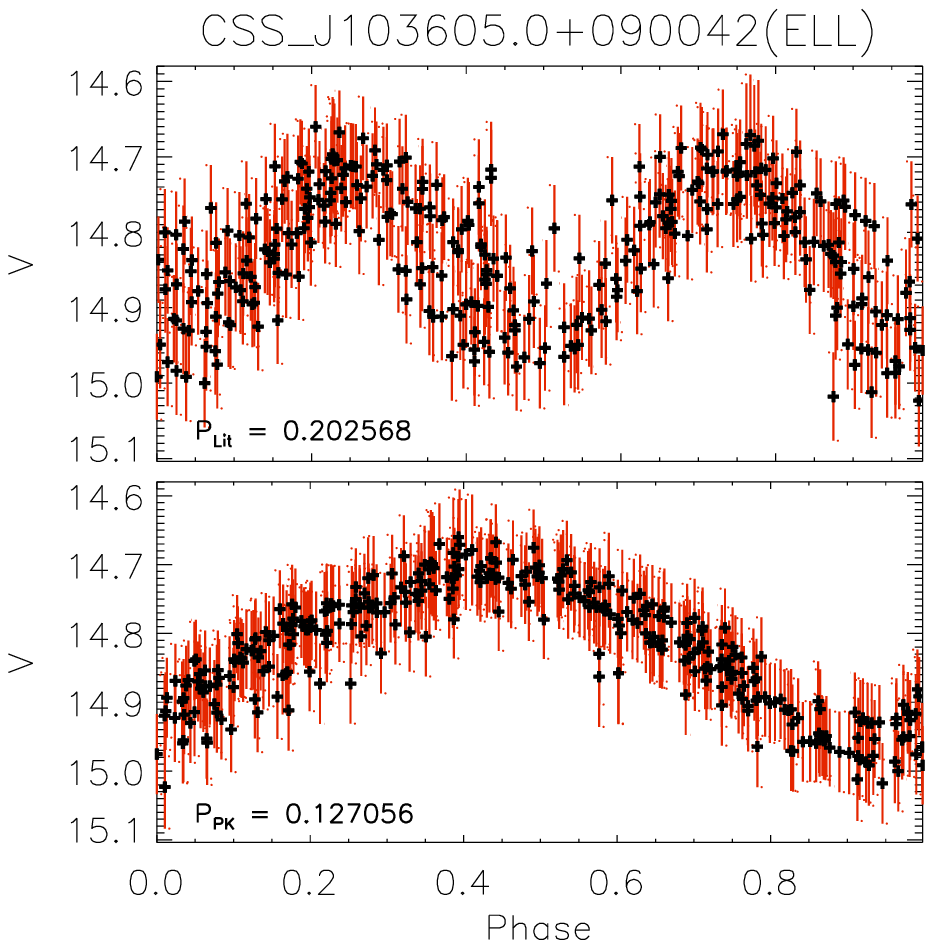}
  \includegraphics[width=0.32\textwidth,height=0.25\textwidth]{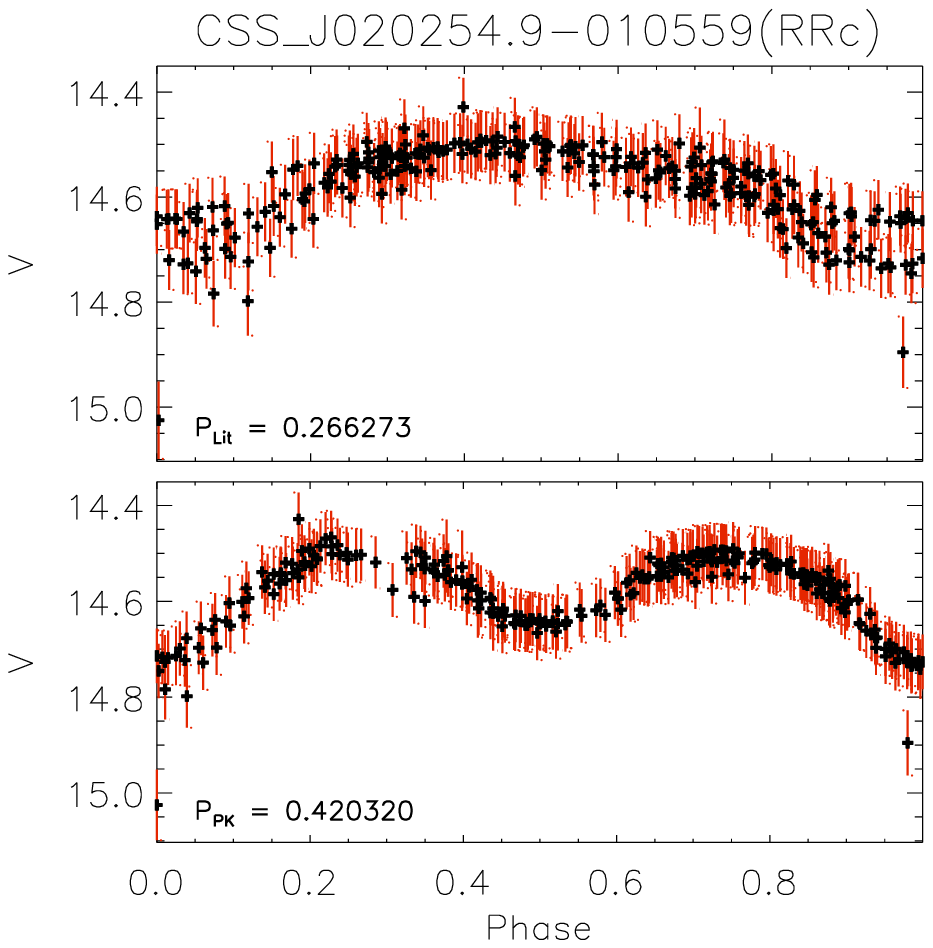}

  \includegraphics[width=0.32\textwidth,height=0.25\textwidth]{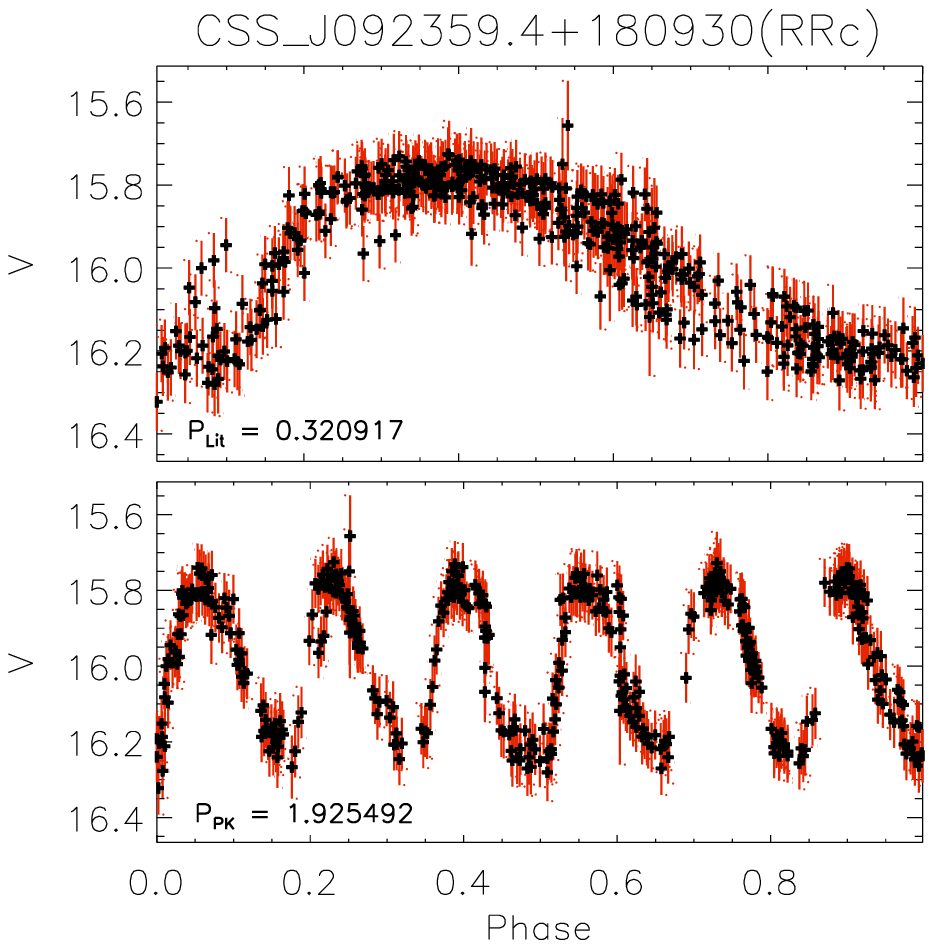}
  \includegraphics[width=0.32\textwidth,height=0.25\textwidth]{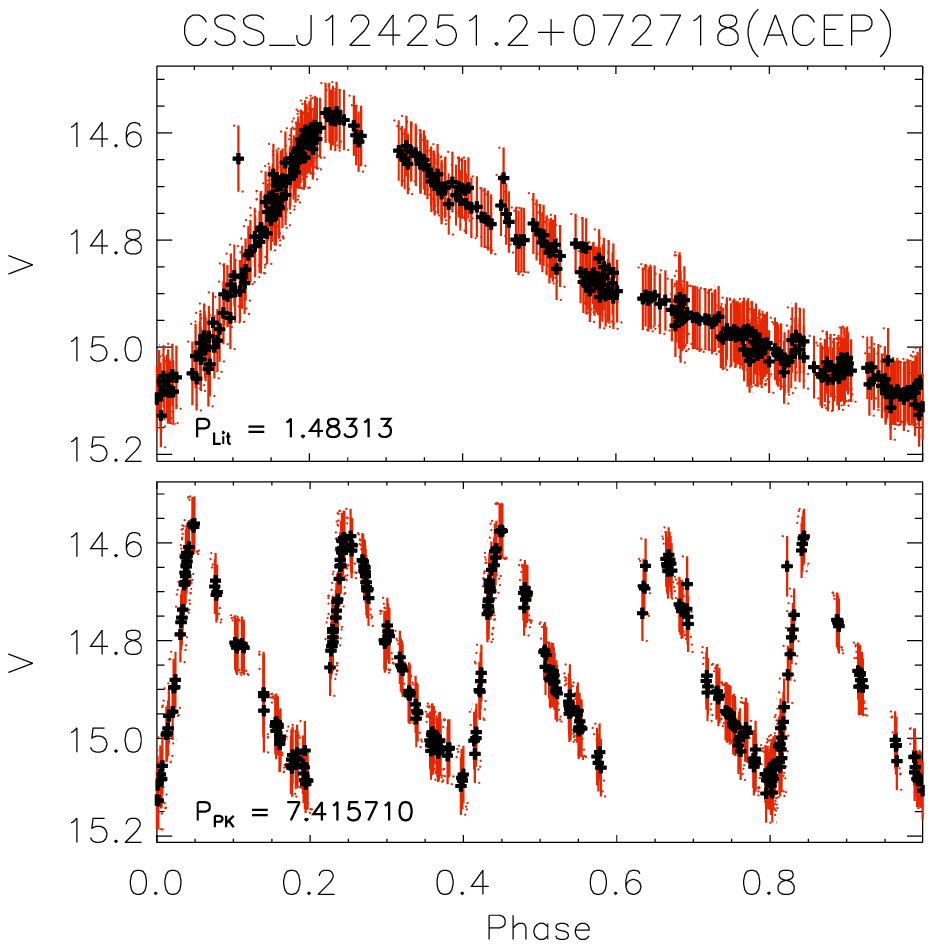}
  \includegraphics[width=0.32\textwidth,height=0.25\textwidth]{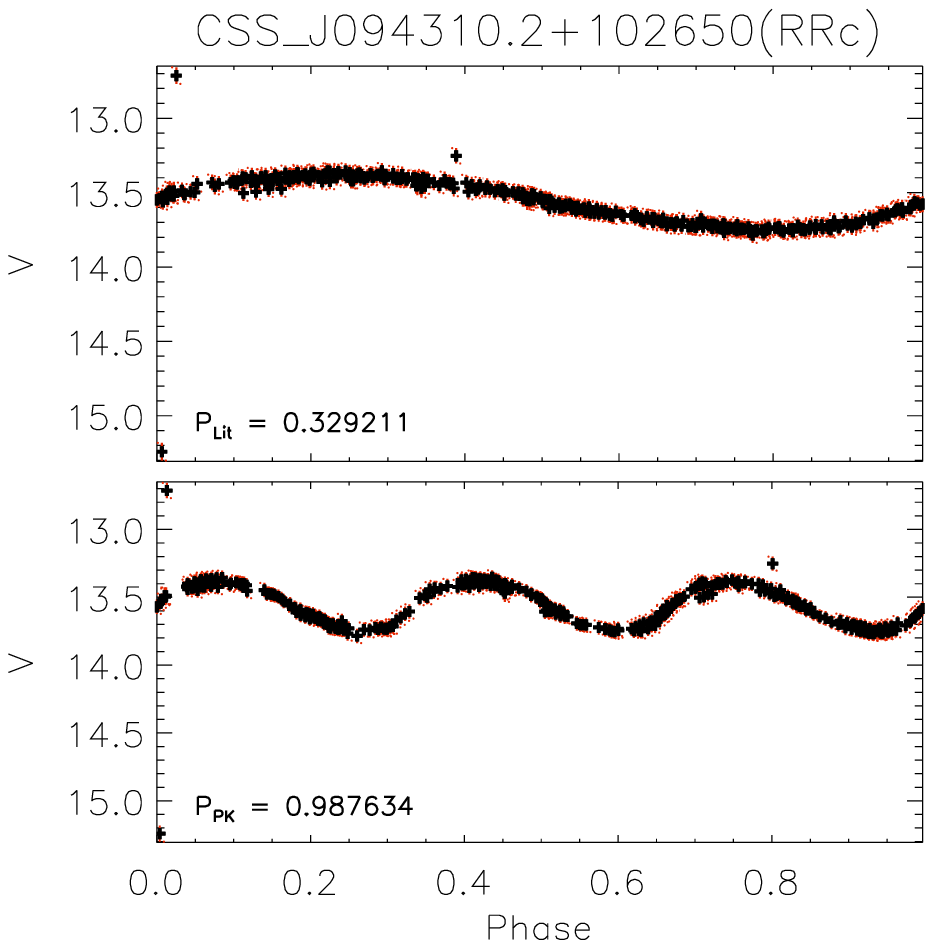}

  \includegraphics[width=0.32\textwidth,height=0.25\textwidth]{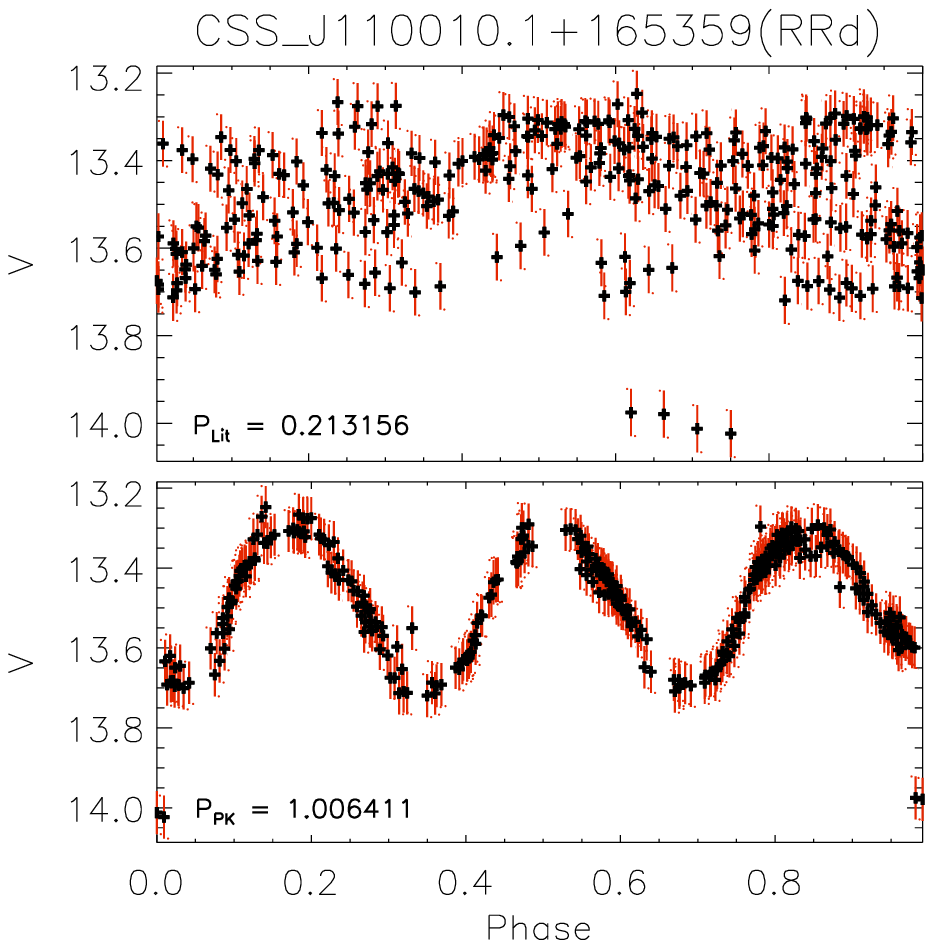}
  \includegraphics[width=0.32\textwidth,height=0.25\textwidth]{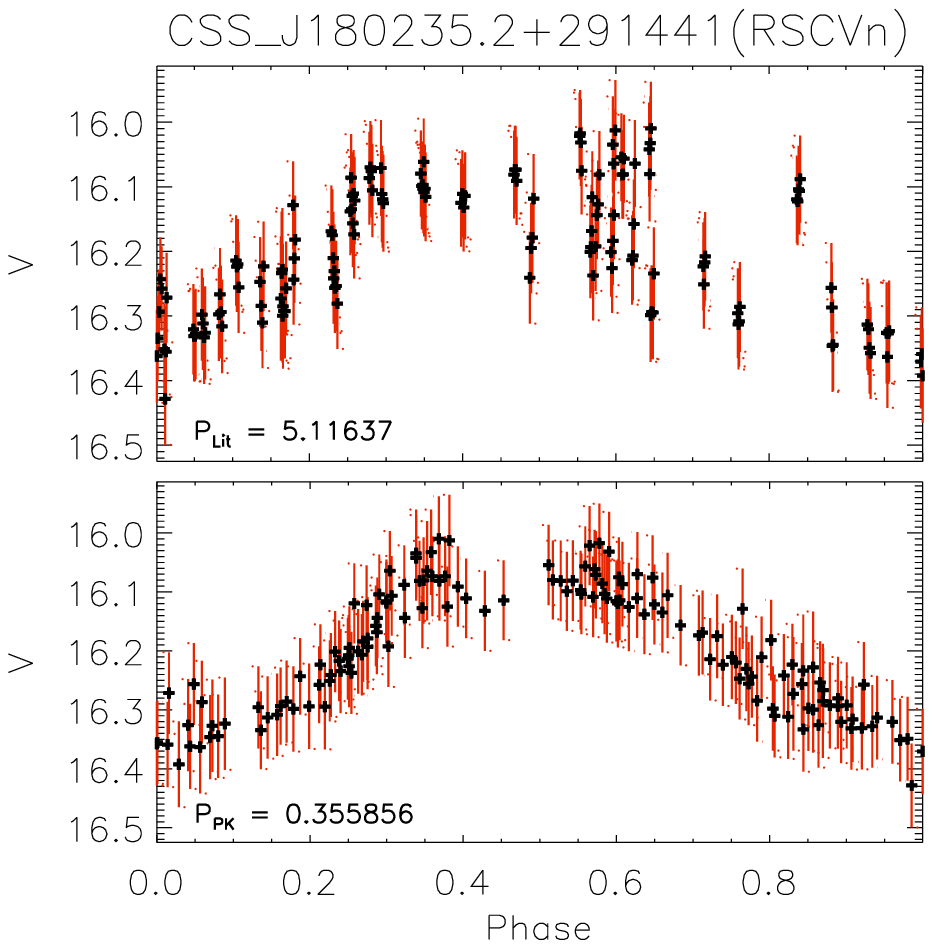}
  \includegraphics[width=0.32\textwidth,height=0.25\textwidth]{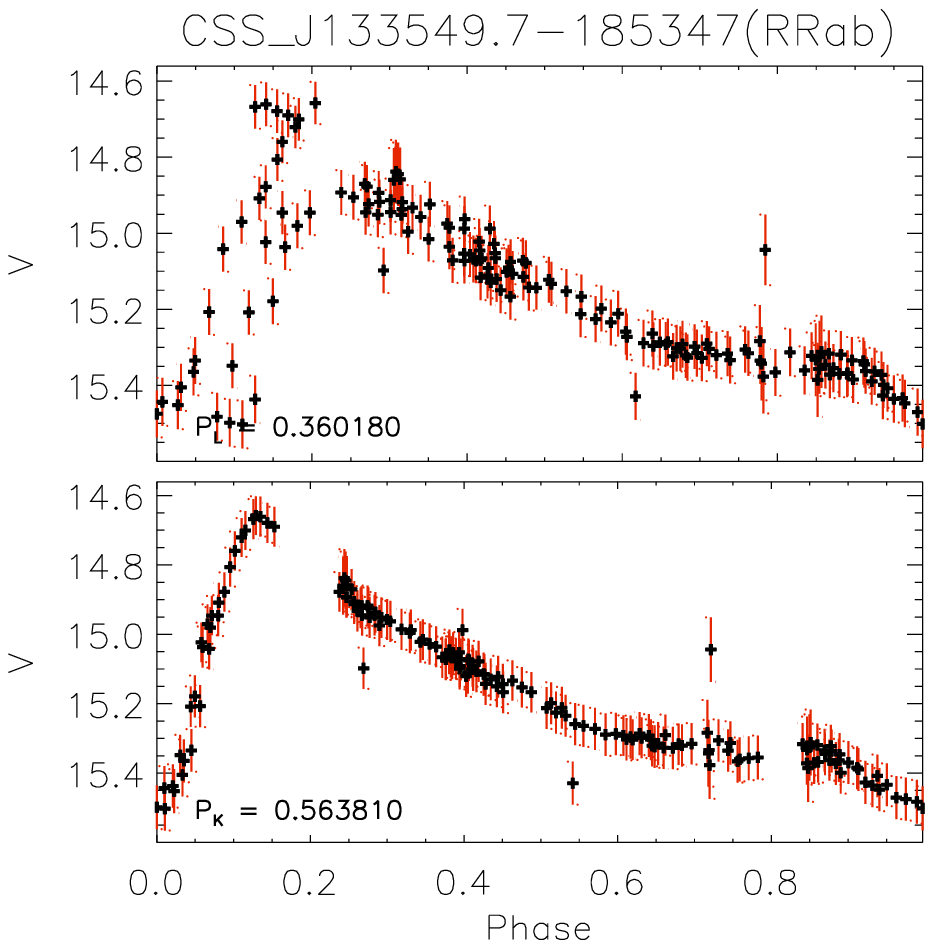}
  
  \caption{Phase diagrams for \textit{CVSC1} stars considering the published 
  variability period ($P_{Lit}$) and that one found by $PK^{(2)}$ method
  ($P_{PK}$).  The star name is shown on the top of each diagram while the periods are in the bottom left corners. }
  \label{fig_lcCatalina}
\end{figure*}

On the other hand, about $\sim8\%$ of the Catalina periods do not
correspond to our periods. We also made a quick visual inspection of the
phased data using the periods found by Catalina and those found by us. As a
result, we verify that the large majority (more that $\sim70\%$) of phase 
diagrams of this subsample produce smoother phase diagrams using our periods
than those found using Catalina periods (see third row of panels of Fig. 
\ref{fig_lcCatalina}). Indeed, this assumes that the true or main variability
period should be that one which produces the smoothest phase diagram.

In summary, the $PK^{(s)}$ and $PL^{(s)}$ methods can be used as a new tool to
find periodic signals. In fact, they are more efficient than all methods tested 
if high SNR data are considered. As a rule, the new approach can be used in the 
same fashion as other period finding methods. One should be aware that, the 
lower efficiency rate for small SNR, probable bias for longer periods, 
and the multiple-periods given by $PK^{(s)}$must be taken into account.

\subsection{Cautionary notes on period searching}\label{sec_pcaution}

The main variability period is assumed to be that one which provides the
smoothest phase diagram. Indeed, the periodicity of many signals in 
oversampled data like CoRoT and Kepler light curves
\citep[e.g.][]{Paz-Chinchon-2015, FerreiraLopes-2015mgiant} can be easily
identified by looking directly at the light curve. This may not include low SNR
multi-periodic signals. On the other hand, the signals in undersampled data
can only can be identified by eye using phase diagrams. In both cases, the
phase diagram should be smooth at the main variability period.  However, more
than one period can lead to smooth phase diagrams. In fact, due to the nature 
of the analysis of big-datasets it is highly likely that some observational 
biases exist or that pathological cases arise where the combination of random 
or correlated errors, nearby sources, mimic expected variations. Therefore, 
additional information must be put together to solve this puzzle. For instance,
photometric colours, amplitudes, nearby saturated sources, crowded sky
regions, distances, and other information are crucial to confirm the period
reliably.

All configurations that produce smooth phase diagrams return the peaks in the
$PK^{(s)}$ method. The current approach  was designed to find the main
variability period from the viewpoint of correlation. Indeed, the harmonic 
periods also provide peaks in the periodogram since the number of consecutive
measurements that cross the even-mean is only a small increase (see Eq.
\ref{eq_pkmax}), often smaller than random crosses due to noise. Actually, other
signals not related to the main variability period also can lead to smooth phase
diagrams and hence they also have peaks close to $PK^{(s)}_{(max)}$. Moreover, incorrect periods also can be obtained if
all configurations that lead to smoother phase diagrams are not addressed.

The $PK^{(s)}$ method is a useful tool to find all periods that lead 
to smooth phase diagrams. Other methods, e.g. the string length,
or PDM method, or the fitting of truncated Fourier series also lead to smooth
phase diagrams. For completeness, the most prominent peaks should be examined
to evaluate the best candidate for the main variability signal. The best  period
can be assumed to be the one that leads to the smallest $\chi^2$ of a model
computed from the phase diagram \citep[e.g.][]{Drake-2014,
FerreiraLopes-2015wfcam,Torrealba-2015}. Figure \ref{fig_lcCatalina} shows five 
cases from \textit{CVSC1} where topics discussed here are a hindrance. In each 
row of panels are presented some examples as follows:

\begin{itemize}
   \item \textbf{First row of panels:} stars where the  $PK^{(2)}$  method does
   not identify the correct variability period. In these cases, an examination
   of the phase diagrams for the other peaks in $PK^{(2)}$ may help to 
   find the correct value.

   \item \textbf{Second row of panels:} stars for which a smooth phase diagram 
   is not clearly defined by either \textit{CVSCI} or the $PK^{(2)}$ method.
   Therefore, both the $P_{Lit}$ and $P_{PK}$ estimate may be wrong.
   Indeed, \citealt[][]{Drake-2014} use other criteria to define the period 
   reliably. However, this analysis is hindered if other information besides
   of light curve are not available.
   
   \item \textbf{Third row of panels:} both the $P_{Lit}$ and $P_{PK}$
   estimates produce smooth phase diagrams. However, they are not sub-harmonics of one
   another. This means that both periods are sub-harmonic of the main
   variability period or one of them is incorrect. Indeed, these systems 
   also might be a complex systems with multiple periodicities, e.g. an
   eclipsing binary where one of component is a pulsating star. These
   examples  illustrate that the criterion of having a smooth phase diagram per 
   se is not enough to define the variability period.
    
   \item \textbf{Fourth row of panels:} the variability period found by
   $PK^{(2)}$  method is an overtone (greater than 2) of the variability
   period.  Therefore, it indicates that the efficiency rate discussed in Sect.
   \ref{sec_erate} is better if higher sub-harmonics are considered.
   
   \item \textbf{Last row of panels:} stars where $P_{Lit}$ is wrong or
   inaccurate. $P_{PK}$ returns smoother phase diagrams than
   those using $P_{Lit}$. Indeed, the $PK^{(s)}$ period for
   \textit{CSS\_J110010.1+165359} appears to be a sub-harmonic of the true
   variability period. The wrong period determination can result in
   a misclassification since many of parameters used for classification are
   derived from the variability period.

\end{itemize}

The \textit{WVSC1} stars have similar features to those discussed using 
\textit{CVSC1} stars. A quick visual inspection was performed to support our
remarks. A few stars look like those found on the last row of panels shown in 
Fig. \ref{fig_lcCatalina} in both samples. Indeed, the main goal of this work 
is to provide a new way to find and analyse variability time-series.

\section{Conclusions}\label{sec_conclusions}

Two new ways to search variability periods are proposed. These methods are not derived from any previous period finding method. The $PK^{(s)}$ method is characterized by presenting ordinates in the range $0$ to $1$, does not have  a strong dependence on the amplitude of the signal, and also has an analytical equation to determine the $\rm FFN$. Moreover, the weight of outliers is reduced since the method only considers the signs of the correlation signal.
These are unique features that allow us to determine a universal false alarm probability, i.e., the cut-off values that can be applied to any time-series, where it mainly
depends on the SNR of the light curve. In contrast, the $PL^{(s)}$ method uses  the correlation values and provides complementary information about the variability period.

The $PK^{(s)}$ and $PL^{(s)}$ methods were compared with the LSG, PDM, and  SLM methods from real and simulated data having single and multi-wavelength data. As result, the efficiency rate found for LSG and PDM methods are better than all other methods for sub-samples having low ($<3$) or high ($>3$) SNR data. On the other hand, $PK^{(s)}$ and $PL^{(s)}$ efficiency is similar to that found for SLM method for data in both constraints. As expected, the accuracy of all methods is increased for data having high SNR.

In fact, the statistics considered in this paper are unlikely to be  useful for data with multiple periodicities. The current methods were recent applied in the entire data of VVV survey \citep[][]{FerreiraLopes-2020} from where
the periods estimated from five period finding method can be found. This paper is the second of this series about period search methods. Our next paper  will provide our summary of recommendation to reduce running time and  improve the periodicity search on big-data sets.

\section{Data and Materials}

The data underlying this article are available in the Catalina repository\footnote{\url{http://nesssi.cacr.caltech.edu/DataRelease/}} and  in the WFCAM Science Archive - WSA\footnote{\url{http://wsa.roe.ac.uk/}}. A friendly version of the data also can be shared on reasonable request to the corresponding author.

\section*{Acknowledgements}

C. E. F. L. acknowledges a  post-doctoral fellowship from the CNPq. N. J. G. C. acknowledges support from the UK Science and Technology Facilities Council. The authors thank to MCTIC/FINEP (CT-INFRA grant 0112052700) and the Embrace Space Weather Program for the computing facilities at INPE.

\bibliographystyle{mnras}
\bibliography{mylib_1611.bib}

\bsp	
\label{lastpage}
\end{document}